\documentstyle[aps,preprint,tighten,floats,psfig]{revtex}

\begin{document}
\draft
\preprint{\vbox{Submitted to Phys. Rev. D \hfill TRI-PP-96-44\\
                    \mbox{}          \hfill UW-DOE/ER/40427-34-N96}}

 \title{New QCD Sum Rules \\for\\ Nucleon Axial Vector Coupling Constants}
\author{
Frank X. Lee}
\address{
TRIUMF, 4004 Wesbrook Mall
Vancouver, BC, Canada V6T 2A3 \\
\mbox{and}\\
Nuclear Physics Laboratory,
Department of Physics,\\
University of Colorado,
 Boulder, CO 80309-0446, USA}
\author{
Derek B. Leinweber}
\address{
Department of Physics,
University of Washington,
Seattle, WA  98195, USA}
\author{
Xuemin Jin}
\address{
TRIUMF, 4004 Wesbrook Mall,
Vancouver, BC, Canada V6T 2A3 \\
\mbox{and}\\
Center for Theoretical Physics,
Laboratory for Nuclear Science
and Department of Physics,
Massachusetts Institute of Technology,
Cambridge, MA 02139, USA}
\maketitle

\begin{abstract}
Two new sets of QCD sum rules for the nucleon axial coupling constants
are derived using the external-field technique and generalized
interpolating fields.  An in-depth study of the predicative ability of
these sum rules is carried out using a Monte-Carlo based uncertainty
analysis.  The results show that the standard implementation of the
QCD sum rule method has only marginal predicative power for the
nucleon axial coupling constants, as the relative errors are large.
The errors range from approximately 50 to 100\% compared to the
nucleon mass obtained from the same method, which has only 10\% to
25\% error.  The origin of the large errors is examined.
Previous analyses of these coupling constants are based on sum rules
that have poor OPE convergence and large continuum contributions.
Preferred sum rules are identified and their predictions are obtained.
We also investigate the new sum rules with an alternative treatment of
the problematic transitions which are not exponentially suppressed in
the standard treatment.  The new treatment provides exponential
suppression of their contributions relative to the ground state.
Implications for other nucleon current matrix elements are also
discussed.
\end{abstract}


\section{Introduction}

Nucleon matrix elements of axial-vector currents at zero momentum
transfer are characterized by the axial-vector coupling constants (or
axial charges): isovector $g_A$, octet $g^8_A$, isoscalar $g^s_A$, and
flavor singlet $g^0_A$.  In terms of the quark spin content of the
nucleon, one may express these coupling constants as:
\begin{eqnarray}
g_A& = &\Delta u - \Delta d,
\nonumber \\ 
g^8_A& = &\Delta u + \Delta d - 2 \Delta s,
\nonumber \\
g^s_A& = &\Delta u + \Delta d,
\nonumber \\
g^0_A& = &\Delta u + \Delta d + \Delta s,
\label{quark}
\end{eqnarray}
where $\Delta q$ is defined by $\langle ps| \bar{q}\gamma_\mu\gamma_5
q|ps\rangle \equiv \Delta q\,\bar{u}(p,s) \gamma_\mu\gamma_5 u(p,s)$
and $u(p,s)$ is a Dirac nucleon spinor.  Knowledge of any three of the
axial charges completely determines the quark spin content of the
nucleon.  Experimentally, $g_A\approx 1.26$ from neutron beta
decay~\cite{Mont94}, $g^8_A\approx 0.60$~\cite{Close93}, $g^s_A\approx
0.40$ from the EMC data together with strange-baryon
decays~\cite{Ash88}, $g^0_A\approx 0.22$ from SMC~\cite{Adams94} and
0.27 from E143~\cite{Abe95}.  These coupling constants reveal
important information on the non-perturbative and non-valence
structure of the nucleon.  Obviously, understanding these quantities
from QCD, the underlying theory of strong interactions, is an
important theoretical issue.  Within the framework of the QCD sum rule
approach~\cite{SVZ79}, these couplings have been previously studied in
Refs.~\cite{Bely84,Chiu85,Gupta89,Henley92,Bely85,Ioffe92}.  In the 
discussions to follow we will use $g_A$ in a generic sense to mean all 
of these couplings.

Recently, a new Monte-Carlo based uncertainty analysis was introduced
to quantitatively determine the predictive ability of QCD sum rules
\cite{Derek96}.  A comprehensive analysis of ground state
$\rho$-meson and nucleon spectral properties was performed.  Many of
the findings contradicted the conventional wisdom of both
practitioners and skeptics alike.  In particular, careful
consideration of opertor product expansion (OPE) convergence and 
ground state dominance revealed that the nucleon sum rule traditionally 
favored in the QCD sum rule approach is invalid.  Furthermore, it was 
found that the nucleon interpolating field advocated by Ioffe is not the 
optimal choice.

In this work, we will apply the Monte-Carlo method to the analysis of
$g_A$.  Our goal is to use this new analysis tool to determine the
predicative ability of the QCD sum rule approach for $g_A$.  This is
the first application of such Monte-Carlo based analysis to a
three-point function.  Since previous studies of $g_A$ are based on
the Ioffe current and the invalid nucleon-mass sum rule, it is
important to re-investigate $g_A$ using alternate nucleon
interpolating fields.  To serve our purpose, two new sets of QCD sum
rules are derived.  One set is derived from a generalized correlator
of spin-1/2 interpolating fields, and the other with a mixed
correlator of spin-1/2 and spin-3/2 fields.  We will examine all of
the new sum rules and compare their performances.

The organization is as follows.  Sec.~\ref{external} sets up the basic
ingredients for calculating $g_A$ using the external field method.  In
Sec.~\ref{spin12}, the QCD sum rules for the spin-1/2 correlator are
presented and compared, wherever possible, with those obtained in
previous works.  In Sec.~\ref{mixed}, the QCD sum rules for the mixed
correlator are given, along with a discussion of the phenomenological
representation.
Results of the Monte-Carlo based sum rule analysis are presented in
Sec.~\ref{results}. In Sec.~\ref{pc} we give an alternative treatment
of the off-diagonal transitions and compare with the standard
treatment.  The summary and conclusions are given in Sec.~\ref{con}.

\section{Calculation of \pmb{$g_A$} using the External Field Method}
\label{external}

The external field approach proceeds by adding an axial-vector
coupling term to the QCD Lagrangian:
\begin{equation}
\Delta {\cal L}=-\sum_q\, g_q \bar{q}\,\hat{Z}\gamma_5\,q,
\end{equation}
and considering the two-point current correlation function 
in the presence of a constant axial vector external field: 
\begin{equation}
\Pi(p)=i\int d^4x\; e^{ip\cdot x}\langle\Omega\,|\,
T\{\;\eta(x)\,\bar{\eta}(0)\;\}\,|\,\Omega\rangle_Z,
\end{equation}
where $|\Omega\rangle$ denotes the QCD vacuum,
$Z$ is the axial vector external field and the hat notation denotes
$\hat{Z}=Z^\alpha\,\gamma_\alpha$.  The coupling $g_q$ keeps track of
the quark flavor in the external field.  For example, if the external
current is $J^5_\mu=\bar{u}\gamma_\mu\gamma_5 u -
\bar{d}\gamma_\mu\gamma_5 d $, then $g_u=-g_d=1$, $g_s=0$.  In this  
way, we can obtain sum rules for all the couplings in the same
calculation.  To first order in the external field, $\Pi(p)$ is
written as:
\begin{equation}
\Pi(p)=\Pi^{(0)}(p)+Z^\alpha\Pi^{(1)}_\alpha(p)+\cdots.
\end{equation}
The axial vector coupling constant $g_A$ is then extracted from the
sum rules for the linear response $\Pi^{(1)}_\alpha(p)$.  The external
field couples directly to the quarks in the nucleon interpolating
fields and polarizes the QCD vacuum. The latter can be described by
the introduction of vacuum susceptibilities.
The external field formalism is equivalent to the direct three-point
function approach~\cite{Burk96}; the resulting sum rules are
identical.

\subsection{Nucleon Interpolating Fields}

The most general spin-1/2 nucleon interpolating current without
derivatives can be written as
\begin{equation}
\eta_{\scriptscriptstyle 1/2}=\eta_1+\beta\,\eta_2,
\label{curr12}
\end{equation}
where $\beta$ is a real parameter and 
\begin{equation}
\eta_1(x)=\epsilon^{abc}\left(u^{aT}(x)C\gamma_5 d^b(x)\right)u^c(x),
\end{equation}
\begin{equation}
\eta_2(x)=\epsilon^{abc}\left(u^{aT}(x)Cd^b(x)\right)\gamma_5 u^c(x).
\end{equation}
The traditional current advocated by Ioffe in QCD sum rule
calculations may be recovered by setting $\beta=-1$ and multiplying an
overall factor of $-2$:
\begin{equation}
\eta_{\scriptscriptstyle N}=2(\eta_2-\eta_1)=
\epsilon^{abc}\left(u^{aT}C\gamma_\mu
d^b\right)\gamma_5 \gamma^\mu u^c
\end{equation}
%
However, we consider the generalized interpolator of (\ref{curr12})
and select $\beta$ to provide optimal sum rules \cite{leinweber95}.

In this work, we also consider the mixed correlator of the generalized
case of Eq.~(\ref{curr12}):
\begin{equation}
\eta_{\scriptscriptstyle \mu,1/2}=\gamma_\mu\gamma_5\,
\eta_{\scriptscriptstyle 1/2},
\label{curr12g}
\end{equation}
with a spin-3/2 current which is also known to couple to the nucleon 
through its spin-1/2 component:
\begin{equation}
\eta_{\scriptscriptstyle \mu,3/2}=\epsilon^{abc}\;[\;
(u^{aT}C\sigma_{\rho\lambda} d^b)\sigma^{\rho\lambda}
\gamma_\mu u^c
-(u^{aT}C\sigma_{\rho\lambda} u^b)\sigma^{\rho\lambda}
\gamma_\mu d^c\;].
\label{curr32}
\end{equation}
The consideration of the mixed correlator is motivated by its success
in nucleon mass sum rules~\cite{Derek96,Furn95}.  There it is
demonstrated that the sum rules from the mixed correlator have greater
overlap with the ground state pole relative to the continuum
contributions, have broader valid Borel regimes, and provide more
stable estimates of spectral properties.  It would be interesting to
see if similar advantages can be gained here.

The matrix elements of the currents between the nucleon and the 
vacuum are defined as 
\begin{equation}
\langle\Omega|\eta_{\scriptscriptstyle 1/2}|N\rangle=\lambda_{1/2}\; u(p)
\end{equation}
\begin{equation}
\langle\Omega|\eta_{\scriptscriptstyle \mu,3/2}|N\rangle
=\lambda_{3/2}\left({4p_\mu\over M_N}+
\gamma_\mu\right) \gamma_5  u(p).
\end{equation}
where $u(p)$ denotes the Dirac spinor of the nucleon and $\lambda$
describes the coupling strength of the currents to a nucleon state.
We use the Dirac spinor normalization $\bar{u}(p)u(p)=2M_N$.

\subsection{Quark Propagator in the External Field}

The calculation of the correlation function in the OPE requires the
fully interacting quark propagator in coordinate space in the presence
of the external axial-vector field.  In the fixed-point gauge, the
propagator up to order $x^2$ and $Z$ is given
by~\cite{Bely84,Chiu85,Henley92,Derek96}
\begin{eqnarray}
& &S^{ab}_q(x,0;Z) \equiv 
\langle\Omega\,|\,T\{q^a(x)\bar{q}^b(0)\}\,|\,\Omega\rangle_Z
\nonumber \\ & &=
 {i \over 2\pi^2}\,{\hat{x}\over x^4} \,\delta^{ab}
\nonumber \\ & &
-\,{1\over 12}\langle\bar{q}q\rangle \delta^{ab}
\nonumber \\ & &
+\,{i\over 32\pi^2}\,
(g_cG^n_{\alpha\beta})\,{\hat{x}\,\sigma^{\alpha\beta}
+\sigma^{\alpha\beta}\,\hat{x}\over x^2}\,{\lambda^{nab}\over 2}
\nonumber \\ & &
+\,{1\over 48}\,{i\over 32\pi^2}\,
(g_cG^n_{\alpha\beta})\,{\hat{x}\,\sigma^{\alpha\beta}
+\sigma^{\alpha\beta}\,\hat{x}\over x^2}\,{\lambda^{nab}\over 2}
\nonumber \\ & &
-\,{1\over 192}\langle\bar{q}g_c\sigma\cdot Gq\rangle
\sigma^{\alpha\beta}\;{\lambda^{ab}\over 2}
\nonumber \\ & &
+\,{1\over 192}\langle\bar{q}g_c\sigma\cdot Gq\rangle\,x^2\,\delta^{ab}
\nonumber \\ & &
-\, {g_q\over 2\pi^2}\,{(x\cdot Z)\,\hat{x}\,\gamma_5\over x^4}\,
 \delta^{ab}
\nonumber \\ & &
+\, {ig_q\over 36}\,\langle\bar{q}q\rangle\,(\hat{x}\hat{Z}-x\cdot Z)
\gamma_5 \,\delta^{ab}
\nonumber \\ & &
+\, {g_q\over 12}\,\chi_v\,\langle\bar{q}q\rangle\,\hat{Z}\,
\gamma_5\,\delta^{ab}
\nonumber \\ & &
+\, {g_q\over 216}\,\kappa_v\,\langle\bar{q}q\rangle\,
\left({5\over 2}x^2\hat{Z}-
x\cdot Z\,\hat{x}\right)\gamma_5\, \delta^{ab}
\nonumber \\ & &
-\,{g_q\over 96}\,\kappa_v\,\langle\bar{q}q\rangle\,
\,(\hat{Z}\,\sigma^{\alpha\beta}
+\sigma^{\alpha\beta}\,\hat{Z})\,\gamma_5\,{\lambda^{nab}\over 2}
\nonumber \\ & &
+ \mbox{higher order terms}.
\label{prop}
\end{eqnarray}
The external-field-induced vacuum susceptibilities $\chi_v$ and
$\kappa_v$ are defined by
\begin{eqnarray}
\langle\bar{q}\gamma_\mu\gamma_5 q\rangle_Z 
&\equiv&
\lim_{Q_\mu\rightarrow 0}\,(iZ_\nu)\,\int d^4x\, e^{iQ\cdot x}\,
\langle \Omega|
T\left[\,\sum_q g_q\, \bar{q}(x)\gamma_\nu\gamma_5 q(x),\,
\bar{q}(0)\gamma_\mu\gamma_5 q(0)\,\right]|\Omega\rangle
\nonumber \\ &\equiv& 
 g_q\,Z_\mu\,\chi_v\,\langle\bar{q}q\rangle,
\end{eqnarray}
and similarly,
\begin{equation}
\langle\bar{q}g_c\tilde{G}_{\mu\nu}\gamma^\nu q\rangle_Z
\equiv g_q\,Z_\mu\,\kappa_v\,\langle\bar{q}q\rangle,
\;\;\;\tilde{G}_{\mu\nu}={1\over2}\epsilon_{\mu\nu\alpha\beta}\,
G^{\alpha\beta}.
\end{equation}
They describe the response of nonperturbative QCD vacuum to
the external field.

\section{QCD sum rules for the spin-1/2 correlator}
\label{spin12}

The calculation of the QCD side proceeds by contracting out 
the quark pairs in the correlation function: 
\begin{equation}
\Pi(p)=i\int d^4x\; e^{ip\cdot x}\langle\Omega\,|\,
T\{\;\eta_{\scriptscriptstyle 1/2}(x)\,
\bar{\eta}_{\scriptscriptstyle 1/2}(0)\;\}\,|\,\Omega\rangle_Z,
\end{equation}
where $\eta_{\scriptscriptstyle 1/2}$ is given in Eq.~(\ref{curr12}),
resulting in the following master formula before Fourier
transformation:
\begin{eqnarray}
& &
\langle\Omega\,|\, T\{\;\eta_{\scriptscriptstyle 1/2}(x)\,
\bar{\eta}_{\scriptscriptstyle 1/2}(0)\;\}\,|\,\Omega\rangle_Z
\nonumber \\ & &
=-\epsilon^{abc}\epsilon^{a^\prime b^\prime c^\prime} \{\;
\beta^2 \gamma_5 S^{aa^\prime}_u \gamma_5
\mbox{Tr}( C {S^{cc^\prime}_d}^T C S^{bb^\prime}_u)
\nonumber \\ & &
+S^{aa^\prime}_u \mbox{Tr}(C {S^{cc^\prime}}^T_d C 
\gamma_5 S^{bb^\prime}_u \gamma_5)
\nonumber \\ & &
+\beta \gamma_5  S^{aa^\prime}_u 
\mbox{Tr}( C {S^{cc^\prime}_u}^T C S^{bb^\prime}_d \gamma_5)
\nonumber \\ & &
+\beta S^{aa^\prime}_u \gamma_5
\mbox{Tr}( C {S^{cc^\prime}_u}^T C \gamma_5 S^{bb^\prime}_d)
\nonumber \\ & &
+ S^{aa^\prime}_u \gamma_5
C {S^{cc^\prime}_d}^T C  \gamma_5 S^{bb^\prime}_u
\nonumber \\ & &
+\beta\gamma_5  S^{aa^\prime}_u \gamma_5
C {S^{cc^\prime}_d}^T C S^{bb^\prime}_u
\nonumber \\ & &
+\beta  S^{aa^\prime}_u
C {S^{cc^\prime}_d}^T C \gamma_5 S^{bb^\prime}_u \gamma_5
\nonumber \\ & &
+\beta^2 \gamma_5  S^{aa^\prime}_u 
C {S^{cc^\prime}_d}^T C S^{bb^\prime}_u \gamma_5\;\},
\label{master11}
\end{eqnarray}
where
\begin{equation}
S^{ab}_q(x,0;Z)\equiv \langle\Omega\,|\,T\{q^a(x)\bar{q}^b(0)\}\,|\,
\Omega\rangle_Z,
\end{equation}
is the quark propagator given in Eq.~(\ref{prop}).  As discussed
above, we are only concerned with the linear response of the
correlator to the external field.  When substituting the quark
propagator into Eq.~(\ref{master11}), we keep terms only to first
order in the external field.
The results after Fourier transform have three distinct Dirac
structures and can be organized by the invariant functions:
\begin{equation}
Z\cdot\Pi^{(1)}(p^2) =
\Pi_1(p^2)\; \hat{Z}\gamma_5
+\Pi_2(p^2)\;Z\cdot p\, \hat{p}\gamma_5 
+\Pi_3(p^2)\;iZ_\mu\sigma^{\mu\nu}p_\nu\gamma_5.
\label{inv11}
\end{equation}
Three sum rules can be derived from the invariant functions.  To save
space, the invariant functions are not written out explicitly. But
they can be easily inferred from the sum rules below.  The sum rules
after Borel transform are as follows.  At structure $\hat{Z}\gamma_5$:
\begin{eqnarray}
& &
\;\;\;{1\over 64}\,[(\beta^2+1)(6g_u+g_d)+2\beta(6g_u+5g_d)]\;
{E_2\,L^{-4/9}\;M^6}
\nonumber \\ & &
-{1\over 24}\,[(\beta^2+1)(9g_u-g_d)+8\beta g_u]\;
{\chi_v a\, E_1\,L^{-4/9}\;M^4}
\nonumber \\ & &
+{1\over 256}\,[(\beta^2+1)(6g_u-g_d)+2\beta(2g_u-g_d)]\;
b\,E_0\, L^{-4/9}\;M^2
\nonumber \\ & &
+{1\over 48}\,[(\beta^2+1)(21g_u+g_d)+2\beta(5g_u-3g_d)]\;
{\kappa_v a\,E_0\,L^{-68/81}\;M^2}
\nonumber \\ & &
-{1\over 72}\,[(\beta^2(18g_u+5g_d)-2\beta(2g_u+3g_d)-14g_u+g_d]\;
{a^2\, L^{4/9}}
\nonumber \\ & &
-{1\over 288}\,[5(\beta^2+1)g_u+\beta(3g_u-g_d)]\;
{\chi_v a\,b \, L^{-4/9}} 
\nonumber \\ & &
=\tilde{\lambda}_{1/2}^2\;\left[\;
g_A\;(1-{2M^2_N\over M^2})\,+\,A\;\right]\;e^{-M_N^2/M^2},
\label{sum1}
\end{eqnarray}
at structure $Z\cdot p\, \hat{p}\gamma_5$:
\begin{eqnarray}
& &
\;\;\;{1\over 64}\,[(\beta^2+1)(6g_u+g_d)+2\beta(6g_u+5g_d)]\;
{E_1\,L^{-4/9}\;M^4}
\nonumber \\ & &
-{1\over 48}\,[(\beta^2+1)g_d+2\beta (2g_u+3g_d)]\;
{\chi_v a\, E_0\,L^{-4/9}\;M^2}
\nonumber \\ & &
+{1\over 256}\,[(\beta^2+1)(6g_u-g_d)+2\beta(2g_u-g_d)]\;
{b\; L^{-4/9}}
\nonumber \\ & &
-{1\over 144}\,[(\beta^2+1)(9g_u+5g_d)+10\beta(g_u+g_d)]\;
{\kappa_v  a\;L^{-68/81}}
\nonumber \\ & &
-{1\over 72}\,[\beta^2(18g_u+g_d)+2\beta (2g_u-3g_d)+(-22g_u+5g_d)]\;
a^2\,L^{4/9}\, {1\over M^2}
\nonumber \\ & &
+{1\over 1152}\,[(\beta^2+1)(2g_u+3g_d)+2\beta g_d]\;
\,\chi_v\,a\,b\, L^{-4/9}\;{1\over M^2}
\nonumber \\ & &
=\tilde{\lambda}_{1/2}^2\;\left[\;
{g_A\over M^2}\,+\,A\;\right]\;e^{-M_N^2/M^2}\ ,
\label{sum2}
\end{eqnarray}
and at structure $iZ_\mu\sigma^{\mu\nu}p_\nu\gamma_5$:
\begin{eqnarray}
& &
-{1\over 48}\,[\beta^2(6g_u+g_d)-2\beta g_d-(6g_u-g_d)]\;
{a\,E_0\,  L^{2/9}\,M^2}
\nonumber \\ & &
+{1\over 24}\,[(\beta^2 g_d-2\beta g_u+2g_u-g_d]\; \chi_v\,a^2
\nonumber \\ & &
-{1\over 432}\,[\beta^2(6g_u+7g_d)-26\beta g_u+20g_u-7g_d]\;
\kappa_v\,a^2\, L^{-32/81}\;{1\over M^2}
\nonumber \\ & &
-{1\over 576}\,[\beta^2(2g_u+7g_d)-6\beta g_u+4g_u-7g_d]\;
\chi_v\,m^2_0\,a^2\,L^{-14/27}\;{1 \over M^2}
\nonumber \\ & &
=\tilde{\lambda}^2_{1/2}\;\left[\;
{g_A\;M_N\over M^2}\,+\,A\;\right]\;e^{-M_N^2/M^2}.
\label{sum3}
\end{eqnarray}
Here and in the following, $a=-(2\pi)^2\,\langle\bar{q}q\rangle$,
$b=\langle g^2_c\, G^2\rangle$, $\langle\bar{q}g_c\sigma\cdot G
q\rangle=-m_0^2\,\langle\bar{q}q\rangle$,
$\tilde{\lambda}_{1/2}=(2\pi)^2\lambda_{1/2}$.  As usual, the
anomalous dimension corrections of the various operators are taken
into account via the factor $L=\left[{\alpha_s(\mu^2)/
\alpha_s(M^2)}\right] =\left[{\ln(M^2/\Lambda_{QCD}^2)/
\ln(\mu^2/\Lambda_{QCD}^2)}\right]$, where $\mu=500$ MeV is the
renormalization scale and $\Lambda_{QCD}$ is the QCD scale
parameter which will be given later.  
The factors $E_n(x)=1-e^{-x}\sum_n{x^n/n!}$ with $x=w^2/
M^2$ account for the excited state contributions, where $w$ is an
effective continuum threshold.  The parameter $A$ is introduced to
account for all contributions from transitions between the nucleon
ground state and the excited states; such a treatment is an approximation
and may lead to errors in the extracted ground state property 
(see Sec.~\ref{ansatz} below).
The continuum threshold and transition strength are {\it a priori}
unknown. So one should bear in mind that they are in principle
different for different sum rules.  We will treat them as parameters
and study their roles in the analysis.

The sum rules for $g_A$ can be obtained by setting $g_u=-g_d=1$ in
Eq.~(\ref{sum1}) to Eq.~(\ref{sum3}), while the sum rules for $g^s_A$,
$g^8_A$, $g^0_A$ can be obtained by setting $g_u=g_d=1$.  Note that
the three axial couplings $g^s_A$, $g^8_A$, $g^0_A$ share the same set
of sum rules. The difference lies in the susceptibilities $\chi_v$ and
$\kappa_v$.

At this point, comparisons can be made with the sum rules obtained in
previous works using the Ioffe current by setting $\beta=-1$ and
$\lambda_{1/2}^2=\lambda^2_N/4$ in Eq.~(\ref{sum1}) to
Eq.~(\ref{sum3}).
For the most part, we find that our sum rules for $g_A$ agree with
those of Refs.~\cite{Bely84,Chiu85,Gupta89,Henley92}, with only a few
exceptions.  There are differences in the anomalous dimension
corrections to the terms involving the mixed condensate.  We use
$-2/27$ for the anomalous dimension of the mixed condensate, while
previous works simply used $0$.  The dimension 7 operator $\chi_v a
b$, which is considered in this work and in Ref.~\cite{Henley92},
differs at structures $\hat{Z}\gamma_5$ and $Z\cdot p\hat{p}\gamma_5$.
Also, the coefficient in front of the dimension 7 operator $\kappa_v
a^2$ at structure $iZ_\mu\sigma^{\mu\nu}p_\nu\gamma_5$ disagrees with
that of Ref.~\cite{Chiu85}.  For the other axial couplings,
comparisons can only be made at structure $Z\cdot p \hat{p}\gamma_5$.
We find that our sum rules agree with those of
Refs.~\cite{Gupta89,Henley92},
but again report corrections for the $\chi_v a b$ term.

\section{QCD sum rules for the mixed correlator}
\label{mixed}

The correlation function we consider is
\begin{equation}
\Pi_{\mu\nu}(p)=i\int d^4x\; e^{ip\cdot x}\langle\Omega\,|\,
T\{\;\eta_{\scriptscriptstyle \mu,1/2}(x)\,
\bar{\eta}_{\scriptscriptstyle \nu,3/2}(0)\;\}\,|\,\Omega\rangle_Z,
\end{equation}
where $\eta_{\scriptscriptstyle \mu,1/2}$ is given in
Eq.~(\ref{curr12g}) and $\eta_{\scriptscriptstyle \mu,3/2}$ is given
in Eq.~(\ref{curr32}).  In the following, we first discuss the
phenomenological representation of the correlation function, then
calculate the QCD side using the OPE.

\subsection{Phenomenological Ansatz for the Mixed Correlator}
\label{ansatz}

The linear response of the correlation function in the external field
can be written as:
\begin{equation}
Z\cdot \Pi^{(1)}_{\mu\nu}(p) 
=(-iZ^\alpha)\; i\,\int d^4x\; e^{ip\cdot x}\langle\Omega\,|\,
T\{\;\eta_{\scriptscriptstyle \mu,1/2}(x)\;
\int d^4y\,J^5_\alpha (y)\;
\bar{\eta}_{\scriptscriptstyle \nu,3/2}(0)\;\}\,|\,\Omega\rangle.
\end{equation}
After inserting two complete sets of intermediate physical states and 
carrying out the integrations, one has:
\begin{eqnarray}
\Pi^{(1)}_{\mu\nu}(p,Z)
&=& Z^\alpha\;
\sum_{BB^\prime}{-1\over (p^2-M_B^2)(p^2-M_{B^\prime}^2)}
\nonumber \\ & &
\sum_{ss^\prime}\langle\Omega\,|\,
\eta_{\scriptscriptstyle \mu,1/2}\;
|\,Bp s\rangle\langle Bp s\,|\,
\,J^5_\alpha \;
|\,B^\prime p s^\prime\rangle\langle B^\prime p s^\prime\,|\,
\bar{\eta}_{\scriptscriptstyle
\nu,3/2}\,|\,\Omega\rangle.
\label{phen}
\end{eqnarray}
The axial current coupling constant enters via the  
nucleon matrix element:
\begin{equation}
\langle Np s\,|\;J^5_\alpha \; |\,N p s\rangle
\equiv\;g_A\; \bar{u}(p,s)\gamma_\alpha\gamma_5 u(p,s).
\end{equation}

To determine the Dirac structure, let us look at 
the ground state contribution to Eq.~(\ref{phen}) after the spin sums:
\begin{eqnarray}
Z\cdot\Pi^{(1)}_{\mu\nu} &=&
{-\lambda_{1/2}\lambda_{3/2}\;g_A\over (p^2-M_N^2)^2} \;
\gamma_\mu\gamma_5\;
(\hat{p}+M_N)\,\hat{Z}\gamma_5\,
(\hat{p}+M_N)\,\gamma_5\,\left({4p_\nu\over M_N} +\gamma_\nu\right)  
\nonumber \\ &=&
 {-\lambda_{1/2}\lambda_{3/2}\;g_A\over (p^2-M_N^2)^2} \;[\;
(p^2+M^2_N)\; \gamma_\mu\gamma_5\, \hat{Z} \,\gamma_\nu
+{4(p^2+M_N^2)\over M_N}\; \gamma_\mu\gamma_5 \hat{Z} p_\nu 
-2\;\gamma_\mu\gamma_5\,(Z\cdot p)\hat{p} \gamma_\nu
\nonumber \\ &-&
{8\over M_N}\; \gamma_\mu\gamma_5\,(Z\cdot p)\hat{p} p_\nu
-2M_N\; \gamma_\mu\gamma_5\,(\hat{Z}\hat{p}-Z\cdot p)\,\gamma_\nu
-8\; \gamma_\mu\gamma_5\,(\hat{Z}\hat{p}-Z\cdot p)\, p_\nu
\;].
\end{eqnarray}
We see that there are eight distinct structures from which eight sum
rules can be derived.

The pole structure of the correlation function from Eq.~(\ref{phen})
can be written as
%
\begin{equation}
{\lambda_{1/2}\lambda_{3/2}\;g_A\over (p^2-M_N^2)^2}
\;+\; \sum_{N^*}{C_{\scriptscriptstyle N\leftrightarrow N^*}
\over (p^2-M_N^2)(p^2-M_{N^*}^2)}
\;+\;\sum_{N^*}{D_{\scriptscriptstyle N^*\rightarrow N^*}\over 
(p^2-M_{N^*}^2)^2},
\label{pole}
\end{equation}
where $C_{\scriptscriptstyle N\leftrightarrow N^*}$ and 
$D_{\scriptscriptstyle N\leftrightarrow N^*}$ are constants.
The first term is the ground state double pole, the second term
represents the non-diagonal transitions between the nucleon and the
excited states caused by the external field, and the third term
represents the excited state contributions.  Upon Borel transform, one
has
\begin{eqnarray}
& &
{\lambda_{1/2}\lambda_{3/2}\;g_A\over M^2}\;e^{-M_N^2/M^2}
\;+\;e^{-M_N^2/M^2}\;\left[\sum_{N^*}
{C_{\scriptscriptstyle N\rightarrow N^*}
\over M_N^2-M_{N^*}^2}
\left(1-e^{-(M_{N^*}^2-M_N^2)/M^2}\right)\right]
\nonumber \\ & &
\;+\; \sum_{N^*}\;
{D_{\scriptscriptstyle N^*\rightarrow N^*}\over M_{N^*}^2}
\;e^{-M_{N^*}^2/M^2}.
\label{phen-gen-borel}
\end{eqnarray}
We see that the transitions (second term) give rise to a contribution
that is not exponentially suppressed relative to the ground state
(first term).  The strength of such transitions at each structure is
{\it a priori} unknown and is an additional source of contamination in
the determination of $g_A$ not found in mass sum rules.  The usual
treatment of the transitions is to approximate the quantity in the
square brackets by a constant phenomenological parameter, which is to
be extracted from the sum rule along with the ground state property of
interest. Such an approximation has been adopted in the sum rules
(\ref{sum1}) to (\ref{sum3}).  Here we want to stress that the
transition term is in fact a complicated function of the Borel mass
and the usual approximation alters the curvature of the
phenomenological side and hence introduces errors in the extracted
ground state property.  Later in this work, we will present an
alternative method of treating such transitions, which provides
exponential suppression of the transitions relative to the ground
state contribution.  The pure excited state contributions (the last
term) are exponentially suppressed relative to the ground state and
can be modeled in the usual way by introducing a continuum model and
threshold parameter.

\subsection{Calculation of the QCD side}

The master formula is given by
\begin{eqnarray}
& &
\langle\Omega\,|\, T\{\;\eta_{\scriptscriptstyle \mu,1/2}(x)\,
\bar{\eta}_{\scriptscriptstyle \nu,3/2}(0)\;\}\,|\,\Omega\rangle_Z
\nonumber \\ & &
=
\epsilon^{abc}\epsilon^{a^\prime b^\prime c^\prime}\{\;
\beta \gamma_\mu S^{aa^\prime}_u \gamma_\nu \sigma_{\rho\lambda}
\mbox{Tr}( S^{bb^\prime}_d \sigma^{\rho\lambda} C {S^{cc^\prime}_u}^T C)
\nonumber \\ & &
+\gamma_\mu\gamma_5  S^{aa^\prime}_u \gamma_\nu \sigma_{\rho\lambda}
\mbox{Tr}(\gamma_5 S^{bb^\prime}_d \sigma^{\rho\lambda}
C {S^{cc^\prime}}^T_u C)
\nonumber \\ & &
-2\gamma_\mu\gamma_5  S^{aa^\prime}_u \sigma_{\rho\lambda}
C {S^{cc^\prime}_u}^T C \gamma_5 S^{bb^\prime}_d
\gamma_\nu \sigma^{\rho\lambda}
\nonumber \\ & &
-\gamma_\mu\gamma_5  S^{aa^\prime}_u \sigma_{\rho\lambda}
C {S^{cc^\prime}_d}^T C \gamma_5 S^{bb^\prime}_u
\gamma_\nu \sigma^{\rho\lambda}
\nonumber \\ & &
-2\beta\gamma_\mu  S^{aa^\prime}_u \sigma_{\rho\lambda}
C {S^{cc^\prime}_u}^T C  S^{bb^\prime}_d
\gamma_\nu \sigma^{\rho\lambda}
\nonumber \\ & &
-\beta\gamma_\mu  S^{aa^\prime}_u \sigma_{\rho\lambda}
C {S^{cc^\prime}_d}^T C S^{bb^\prime}_u
\gamma_\nu \sigma^{\rho\lambda}\; \}.
\label{master12}
\end{eqnarray}
The calculation proceeds in the same way as in the spin-1/2 case by
substituting the quark propagator into Eq.~(\ref{master12}), keeping
terms to first order in the external field.
The results after Fourier transform have eight distinct Dirac
structures and can be organized as
\begin{eqnarray}
Z\cdot \Pi_{\mu\nu}^{(1)}(p^2) &=&
 \Pi_{1}(p^2)\; \gamma_\mu\gamma_5 \hat{Z} \gamma_\nu
+\Pi_{2}(p^2)\; \gamma_\mu\gamma_5 \hat{Z} p_\nu
+\Pi_{3}(p^2)\; \gamma_\mu\gamma_5 (Z\cdot p) \gamma_\nu
\nonumber \\ & &
+\Pi_{4}(p^2)\; \gamma_\mu\gamma_5 (Z\cdot p) p_\nu
+\Pi_{5}(p^2)\; \gamma_\mu\gamma_5 \hat{Z} \hat{p} \gamma_\nu
+\Pi_{6}(p^2)\; \gamma_\mu\gamma_5 \hat{Z} \hat{p} p_\nu
\nonumber \\ & &
+\Pi_{7}(p^2)\; \gamma_\mu\gamma_5 (Z\cdot p) \hat{p} \gamma_\nu
+\Pi_{8}(p^2)\; \gamma_\mu\gamma_5 (Z\cdot p) \hat{p} p_\nu.
\label{inv12}
\end{eqnarray}

After Borel transform, eight sum rules are obtained.
They are as follows.
At structure $\gamma_\mu\gamma_5\, \hat{Z} \,\gamma_\nu$:
\begin{eqnarray}
& &
-{(1-\beta)(g_u-g_d)\over 4}\, \chi_v\,a\, E_1\,L^{-4/27}\;M^4
\nonumber \\ & &
-{(1-\beta)(g_u-g_d)\over 18}\,\kappa_v\,a\,E_0\,L^{-44/81}\;M^2
-{(1-\beta)(g_u-g_d)\over 96}\,b\,E_0\,L^{-4/27}\; M^2
\nonumber \\ & &
+{\beta(7g_u+5g_d)+10g_u \over 9}\,a^2\,L^{20/27}
+{(1-\beta)(g_u-g_d)\over 96}\,\chi_v\,a\, b \,L^{-4/27}
\nonumber \\ & &
=\tilde{\lambda}_{1/2}\tilde{\lambda}_{3/2}\;\left[\;
g_A\;(1-{2M_N^2\over M^2})\,+\,A\;\right]\;e^{-M_N^2/M^2},
\label{mix1}
\end{eqnarray}
at structure $\gamma_\mu\gamma_5 \hat{Z} p_\nu$:
\begin{eqnarray}
& &
{3\beta(g_u+g_d)+11g_u-2g_d\over 18}\, a\, E_0\,L^{8/27}\; M^2
\nonumber \\ & &
+{3\beta(5g_u+7g_d)+19g_u-7g_d\over 96}\,m_0^2\,a\,L^{-2/9}
-{\beta(g_u-g_d)+7g_u-g_d\over 6}\,\chi_v\,a^2\,L^{8/27}
\nonumber \\ & &
+{\beta(g_u-7g_d)+19g_u-7g_d\over 144}\,\chi_v\,m^2_0\,a^2\,
L^{-2/9}\;{1\over M^2}
\nonumber \\ & &
+{\beta(g_u+11g_d)-3(8g_u+g_d)\over 54}\,\kappa_v\,a^2\,L^{-8/81}\;
{1\over M^2}
\nonumber \\ & &
=\tilde{\lambda}_{1/2}\tilde{\lambda}_{3/2}\;\left[\;
{g_A\over M_N}\;(1-{2M_N^2\over M^2})\,+\,A\;\right]\;e^{-M_N^2/M^2},
\label{mix2}
\end{eqnarray}
at structure $\gamma_\mu\gamma_5 Z\cdot p \gamma_\nu$:
\begin{eqnarray}
& &
{12\beta(g_u+g_d)+8g_u+g_d\over 18}\, a\, E_0\,L^{8/27}\; M^2
-{\beta(g_u+2g_d)\over 3}\,\chi_v\,a^2\,L^{-8/27}
\nonumber \\ & &
+{7\beta(g_u+2g_d)\over 72}\,\chi_v\,m^2_0\,a^2\,L^{-2/9}\;{1\over M^2}
+{\beta(59g_u+55g_d)+9(5g_u-g_d)\over 108}\,\kappa_v\,a^2\,L^{-8/81}\;
{1\over M^2}
\nonumber \\ & &
=\tilde{\lambda}_{1/2}\tilde{\lambda}_{3/2}\;\left[\;
{g_A\,M_N\over M^2}\,+\,A\;\right]\;e^{-M_N^2/M^2},
\label{mix3}
\end{eqnarray}
at structure $\gamma_\mu\gamma_5 Z\cdot p p_\nu$:
\begin{eqnarray}
& &
{(1-\beta)(g_u-g_d)\over 12}\, \chi_v\,a\,  E_0\,L^{-4/27}\;M^2
\nonumber \\ & &
-{5(1-\beta)(g_u-g_d)\over 36}\,\kappa_v\,a\,L^{-44/81}
-{(1-\beta)(g_u-g_d)\over 96}\,b\,L^{-4/27}
\nonumber \\ & &
+{4\beta(2g_u+g_d)+5g_u+2g_d\over 18}\,a^2\,L^{20/27}\;{1\over M^2}
+{(1-\beta)(g_u-g_d)\over 288}\,\chi_v\,a\,b\,L^{-4/27}\;{1\over M^2}
\nonumber \\ & &
=\tilde{\lambda}_{1/2}\tilde{\lambda}_{3/2}\;\left[\;
{g_A\over M^2}\,+\,A\;\right]\;e^{-M_N^2/M^2},
\label{mix4}
\end{eqnarray}
at structure $\gamma_\mu\gamma_5 \hat{Z}\hat{p} \gamma_\nu$:
\begin{eqnarray}
& &
 {3\beta(g_u+g_d)-g_u+g_d\over 12}\, a\,E_0\,L^{8/27}\; M^2
-{\beta(g_u+5g_d)-7g_u+g_d\over 12}\,\chi_v\,a^2\,L^{8/27}
\nonumber \\ & &
+{\beta(13g_u+35g_d)-19g_u+7g_d\over
288}\,\chi_v\,m^2_0\,a^2\,L^{8/27}\;
{1\over M^2}
\nonumber \\ & &
+{\beta(31g_u+47g_d)-73g_u-5g_d\over 216}\,\kappa_v\,a^2\,L^{-8/81}\;
{1\over M^2}
\nonumber \\ & &
=\tilde{\lambda}_{1/2}\tilde{\lambda}_{3/2}\;\left[\;
{g_A\,M_N\over M^2}\,+\,A\;\right]\;e^{-M_N^2/M^2},
\label{mix5}
\end{eqnarray}
at structure $\gamma_\mu\gamma_5 \hat{Z}\hat{p} p_\nu$:
\begin{eqnarray}
& &
{(1-\beta)(g_u-g_d)\over 24}\, \chi_v\,a\, E_0\,L^{-4/27}\; M^2
-{7(1-\beta)(g_u-g_d)\over 144}\,\kappa_v\,a\,L^{-44/81}
\nonumber \\ & &
+{\beta(g_u-g_d)+8g_u-g_d\over 18}\,a^2\,\,L^{20/27}\;{1\over M^2}
+{(1-\beta)(g_u-g_d)\over 576}\,\chi_v\,a\,b\, L^{-4/27}\;{1\over M^2}
\nonumber \\ & &
=\tilde{\lambda}_{1/2}\tilde{\lambda}_{3/2}\;\left[\;
{g_A\over M^2}\,+\,A\;\right]\;e^{-M_N^2/M^2},
\label{mix6}
\end{eqnarray}
at structure $\gamma_\mu\gamma_5 Z\cdot p \hat{p} \gamma_\nu$:
\begin{eqnarray}
& &
{(1-\beta)(g_u-g_d)\over 12}\, \chi_v\,a\, E_0\,L^{-4/27}\;  M^2
\nonumber \\ & &
-{5(1-\beta)(g_u-g_d)\over 36}\,\kappa_v\,a\,L^{-44/81}
-{(1-\beta)(g_u-g_d)\over 96}\,b\,L^{-4/27}
\nonumber \\ & &
+{2(1+\beta)+3g_u+g_d\over 9}\,a^2\,L^{20/27}\;{1\over M^2}
+{(1-\beta)(g_u-g_d)\over 288}\,\chi_v\,a\,b\;L^{-4/27}\;{1\over M^2}
\nonumber \\ & &
=\tilde{\lambda}_{1/2}\tilde{\lambda}_{3/2}\;\left[\;
{g_A\over M^2}\,+\,A\;\right]\;e^{-M_N^2/M^2},
\label{mix7}
\end{eqnarray}
and at structure $\gamma_\mu\gamma_5 Z\cdot p \hat{p} p_\nu$:
\begin{eqnarray}
& &
{3\beta(g_u+g_d)+11g_u-2g_d\over 18}\, a\,L^{8/27}
-{3\beta(5g_u+7g_d)+19g_u-7g_d\over 96}\,m^2_0\,a\,L^{-2/9}\; 
{1\over M^2}
\nonumber \\ & &
-{3\beta(g_u+g_d)+7g_u-g_d\over 27}\,\kappa_v\,a^2\; L^{-8/81}\;
{1\over M^4}
\nonumber \\ & &
=\tilde{\lambda}_{1/2}\tilde{\lambda}_{3/2}\;\left[\;
{g_A\over M_N\;M^2}\,+\,A\;\right]\;e^{-M_N^2/M^2}.
\label{mix8}
\end{eqnarray}
%

\section{Monte Carlo Sum Rule Analysis}
\label{results}

The reader is referred to Ref.~\cite{Derek96} for a complete
description of the method.  The basic steps are as follows.  One first
generates many sets of randomly-selected, Gaussianly-distributed QCD
parameter sets, from which an uncertainty distribution in the OPE can
be constructed.  Then a $\chi^2$ minimization is applied to the sum
rule by adjusting the phenomenological fit parameters.  This is done
for each QCD parameter set, resulting in distributions for
phenomenological fit parameters, from which errors are derived.
Usually, 100 such configurations are sufficient for getting stable
results. We generally select 1000 which help resolve more subtle
correlations among the QCD parameters and the phenomenological fit
parameters.

The Borel window over which the two sides 
of a sum rule are matched is determined by the following two criteria: 
a) {\em OPE convergence} --- the highest-dimension-operators 
contribute no more than 10\% to the QCD side when $\beta = 0$,
b) {\em ground-state dominance} --- all excited
state contributions (including transitions) are no more than 50\% of the 
phenomenological side. 
Those sum rules which do not have a valid Borel window are considered
unreliable and therefore discarded.  The emphasis here is on exploring
the QCD parameter space via Monte Carlo.  The 10\%-50\% criteria are a
reasonable choice that provide a basis for quantitative analysis.
Reasonable alternatives to the 10\%-50\% criteria are automatically
explored in the Monte-Carlo analysis, as the condensate values and the
continuum threshold change in each sample.

\subsection{QCD Input Parameters}
 
The QCD input parameters and their uncertainty assignments are given
as follows.  The quark condensate is taken as $a=0.52\pm0.05$ GeV$^3$.
A number of recent studies~\cite{Derek96} prefer much larger values
for the gluon condensate than early estimates of $0.47\pm0.2$ GeV$^4$
from charmonium sum rules.  Hence we adopt $b=1.2\pm0.6$ GeV$^4$ with
50\% uncertainty.  The mixed condensate parameter is placed at
$m^2_0=0.72\pm0.08$ GeV$^2$.  Note that the value of the mixed
condensate itself is obtained by multiplying the randomly-selected
$m^2_0$ with the central value of the quark condensate.  Factorization
violation of the four quark operators is parameterized as
$\kappa\langle\bar{q}q\rangle^2$, where we consider $\kappa=2\pm 1$
and $1\leq \kappa \leq 4$. 
The QCD scale parameter $\Lambda_{QCD}$ is restricted to the values
conventionally adopted in the QCD sum rule approach: 
$\Lambda_{QCD}$=0.15$\pm$0.04 GeV, 
and 0.10 GeV $\le \Lambda_{QCD} \le$ 0.20 GeV. 
Variation of $\Lambda_{QCD}$ has little effects on the results.

The external-field-induced vacuum susceptibilities for the isovector 
$g_A$ have been estimated previously~\cite{Bely84,Chiu85,Nov84}. We consider 
$\chi_v a=0.70\pm 0.05$ GeV$^2$ and $\kappa_v a=0.14\pm 0.14$ GeV$^4$.
$\chi_v a$ is related to PCAC which is well known. 
$\kappa_v$ is related to the matrix element 
$\langle 0|\bar{q}g_c\tilde{G}_{\mu\nu}\gamma^\nu q|\pi\rangle_Z$
which is not well determined.
Note that the combinations $\chi_v a$ and $\kappa_v a$ as a whole 
are selected by Monte Carlo.
For the flavor singlet $g^0_A$, they have been estimated 
in Ref.~\cite{Ioffe92}. Here we adopt them with 
50\% and 100\% uncertainties, respectively:
$\chi_v a=0.14\pm 0.07$ GeV$^2$,
$\kappa_v a=0.01\pm 0.01$ GeV$^4$.

For the octet $g^8_A$ and isoscalar $g^s_A$, the susceptibilities are
estimated from the following consideration.  Nucleon matrix elements
$\langle \bar{q}\gamma_\mu\gamma_5 q\rangle$ have a connected
(valence) contribution, as well as a quark loop contribution connected
only by gluons.  Lattice studies~\cite{Liu95} found that the loop
contributions are almost independent of the quark flavors u, d and s.
In the limit of equal loop contributions for u, d and s quarks, both
$g_A$ and $g^8_A$ have the loop contributions cancelled out, leaving
only the valence contributions, as shown in Eq.(\ref{quark}).  Hence,
we assume the susceptibilities are the same for these two couplings.
On the other hand, $g^0_A$ has valence plus three flavors of loop
contributions, and $g^s_A$ has valence plus two flavors of loop
contributions.  From these arguments one can express the
susceptibilities for $g^s_A$ in terms of those for $g_A$ and $g^0_A$.
They are estimated as $\chi_v=0.33\pm 0.16$ and $\kappa_v=0.05\pm
0.05$, where we have assigned 50\% and 100\% uncertainties,
respectively.

\begin{figure}[p]
\centerline{\psfig{file=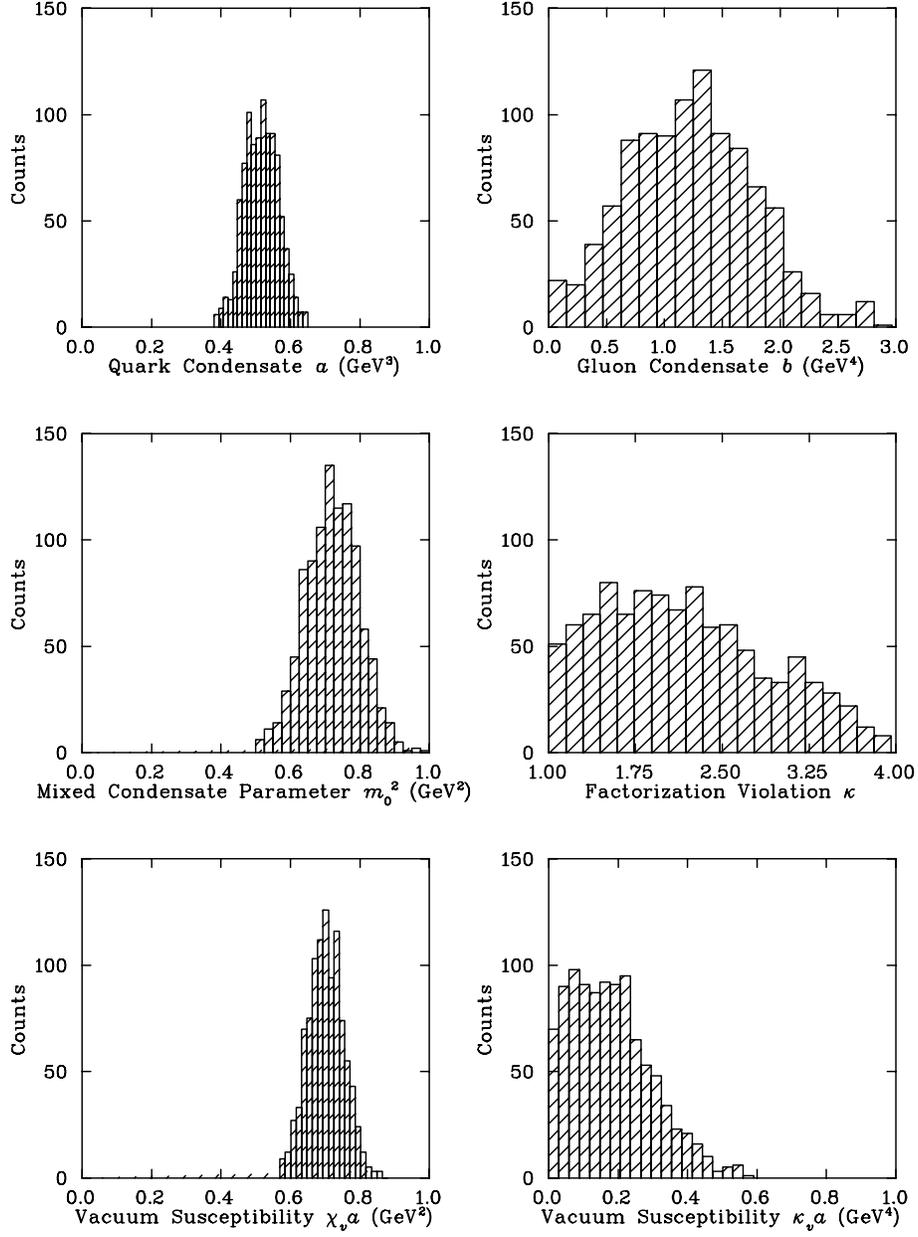,width=12cm}}
\vspace{1cm}
\caption{Distributions for the QCD parameters drawn from a sample of 
1000 sets. The vacuum susceptibilities for $g_A$ are shown here.}
\label{BINQCD}
\end{figure}

These uncertainties are assigned conservatively and in accord with the
state-of-the-art in the literature.  While some may argue that some
values are better known, others may find that the errors are
underestimated.  In any event, one will learn how the uncertainties in
the QCD parameters are mapped into uncertainties in the
phenomenological fit parameters.  Fig.~\ref{BINQCD} shows
distributions for these QCD parameters drawn from a sample of 1000
sets.
%
%

\subsection{Search Procedure}

In principle, one can extract $g_A$ from any of the sum rules
presented earlier.  In practice, however, some sum rules work better
than others. This is because one works with a truncated OPE series,
which may have different convergence properties at different
structures.  By selecting appropriate mixing of the components of the
generalized interpolating fields, one can minimize the overlap with
excited states, and
broaden the regime in Borel space where both sides of the sum rule are
under reasonable control.  In the following, we use the optimal
$\beta=-1.2$ for the spin-1/2 sum rules and $\beta=0$ for the mixed
sum rules, as determined in Ref.~\cite{Derek96}.

The analysis of $g_A$ sum rules requires the corresponding nucleon
mass sum rules for normalization.  They are taken from
Ref.~\cite{Derek96} for the spin-1/2 correlator at structure $1$:
 \begin{eqnarray}
 & &
{7-2\beta-5\beta^2\over 16}\, a\,E_1\;M^4
-{3(1-\beta^2)\over 16}\,m^2_0\,a\,E_0\;L^{-14/27}\; M^2
 \nonumber \\ & &
+{19+10\beta-29\beta^2\over 2^7\,3^2}\,a\,b\
=\tilde{\lambda}^2_{1/2}\,M_N\;e^{-M_N^2/M^2}.
\label{nuc2}
 \end{eqnarray}
Note that we use the nucleon mass sum rule at the chirally-even
structure $1$, rather than the traditionally favored one at the the
chirally-odd structure $\hat{p}$, since the latter was found to be
invalid~\cite{Derek96}.  The optimal fit parameters obtained from
consideration of 1000 QCD parameter sets are
\begin{equation}
M_N=1.17\pm0.26\; \mbox{GeV},\hspace{3mm} 
\tilde{\lambda}^2_{1/2}=0.20\pm0.12\; \mbox{GeV}^6,\hspace{3mm}
w_N=1.53\pm0.41\; \mbox{GeV},\hspace{3mm},
\label{fit1}
\end{equation}
For the mixed correlator, we use the nucleon mass sum rules
at structure $\gamma_\mu\gamma_\nu$ 
\begin{equation}
{1\over 2}\,a\, L^{8/27}\,E_1\;M^4
+{1+3\beta\over 96}\,a\,b\;L^{8/27}
=\tilde{\lambda}_{1/2}\tilde{\lambda}_{3/2}\,M_N\;e^{-M_N^2/M^2},
\label{nucmix2}
\end{equation}
and at structure $\gamma_\mu\hat{p}p_\nu$
\begin{equation}
{1\over 2}\,a\,L^{8/27}\, E_0\;M^2
-{3-\beta\over 16}\,m^2_0\,a\;L^{-2/9}
-{1+3\beta\over 96}\,{a\,b\over M^2}\;L^{8/27}
=\tilde{\lambda}_{1/2}\tilde{\lambda}_{3/2}\,M_N\;e^{-M_N^2/M^2}.
\label{nucmix3}
\end{equation}
A combined analysis of the two sum rules from consideration of 1000
QCD parameter sets gives
\begin{equation}
M_N=0.96\pm0.08\; \mbox{GeV},\hspace{3mm} 
\tilde{\lambda}_{1/2}\tilde{\lambda}_{3/2}=0.41\pm0.14\; \mbox{GeV}^6,
\hspace{3mm} w_N=1.3\pm0.2\; \mbox{GeV},\hspace{3mm}.
\label{fit2}
\end{equation}
Note that the parameters determined from the mixed correlator 
have much smaller uncertainty than from the spin-1/2 correlator.

When performing a Monte-Carlo analysis of a $g_A$ sum rule, we first
fit the corresponding mass sum rule(s) to obtain the nucleon mass
$M_N$, pole residue $\tilde{\lambda}^2$ and the continuum threshold
$w_N$.  Then, the mass and the pole residue are used in the $g_A$ sum
rule where the three remaining parameters including the transition
strength $A$, the continuum threshold $w$, and $g_A$ are optimized.
We impose a physical constraint on $w$ requiring that $w> M_N$,
and discard QCD parameter sets that do not satisfy this criteria.
The above procedure is repeated for each QCD parameter set until a
certain number of sets (typically 1000 sets) are finished.

In doing a full search for the $g_A$ sum rules, we encountered three
scenarios regarding the continuum threshold $w$.  First, $w$ falls
consistently below the nucleon mass, signaling a failure of the sum
rule to resolve the pole from the continuum.  In order to proceed in
this case, the continuum threshold of the mass sum rule is used in the
corresponding $g_A$ sum rule.  Note that this is the assumption made
in previous works analyzing $g_A$.  Second, we are able to get $w>
M_N$ from the search, but the uncertainty on $w$ is
uncharacteristically large, a sign of an unstable fit. The origin of
the large uncertainty is that the search algorithm occasionally
returns very large values for $w$.  Since in the continuum model, the
contributions from large $w$ are exponentially damped out, the
extracted results for $g_A$ and the transition are not seriously
affected. We consider such fits marginally acceptable. One could in
principle impose a cutoff on large $w$ in the search algorithm to
reduce the uncertainty. We choose not to do so, but rather use it as a
performance indicator of the sum rule.  Third, we get both $w> M_N$
and reasonable uncertainty.  This is the best scenario.

\subsection{Results and Discussion}

\begin{figure}[p]
\centerline{\psfig{file=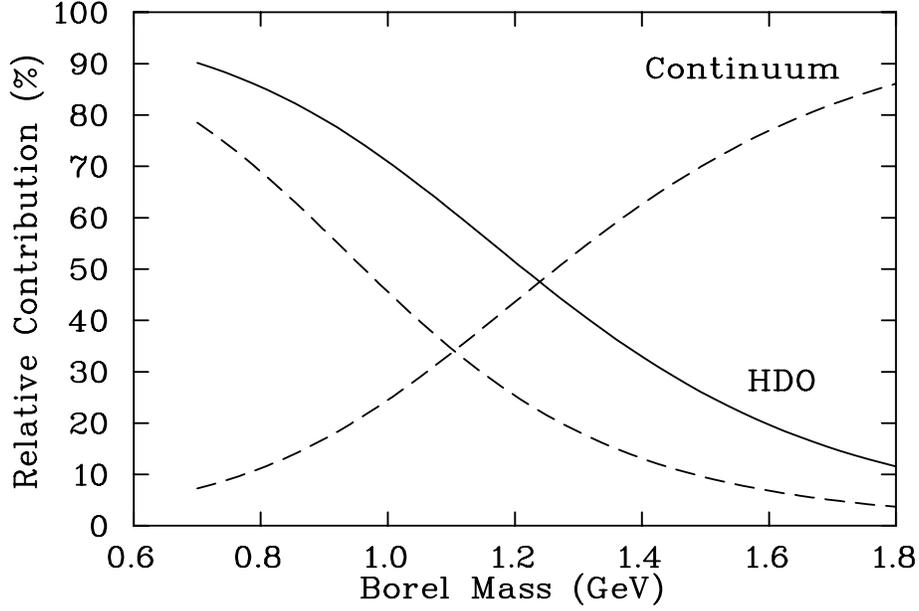,height=8cm,width=12cm,angle=90}}
\vspace{1cm}
\caption{
The HDO contributions of the OPE relative to the sum of terms and the
continuum-plus-transition contributions relative to the total
phenomenological side are displayed as a function of the Borel mass
for sum rule~(\protect\ref{sum2}) at $\beta=-1.2$ (dashed).  The solid
line shows the HDO contributions at $\beta=0$.}
\label{HIDCON7}
\end{figure}

Now we are ready to examine the sum rules.  We find that sum rule
~(\ref{sum2}) at the chirally-odd structure $Z\cdot p\,
\hat{p}\gamma_5$, which has been chosen in previous works, fails to
have a valid Borel window.  Fig.~\ref{HIDCON7} shows the
highest-dimension-operator (HDO) contributions of the QCD side
relative to the sum of terms and the continuum-plus-transition
contributions relative to the total phenomenological side as a
function of the Borel mass.  The former is decreasing with the Borel
mass while the latter is increasing.  Also shown are the HDO
contributions at $\beta=0$, which should be used to determine the
lower limit of the Borel window when the 10\% criteria is
applied~\cite{Derek96}.
%
%
\begin{figure}[p]
\centerline{\psfig{file=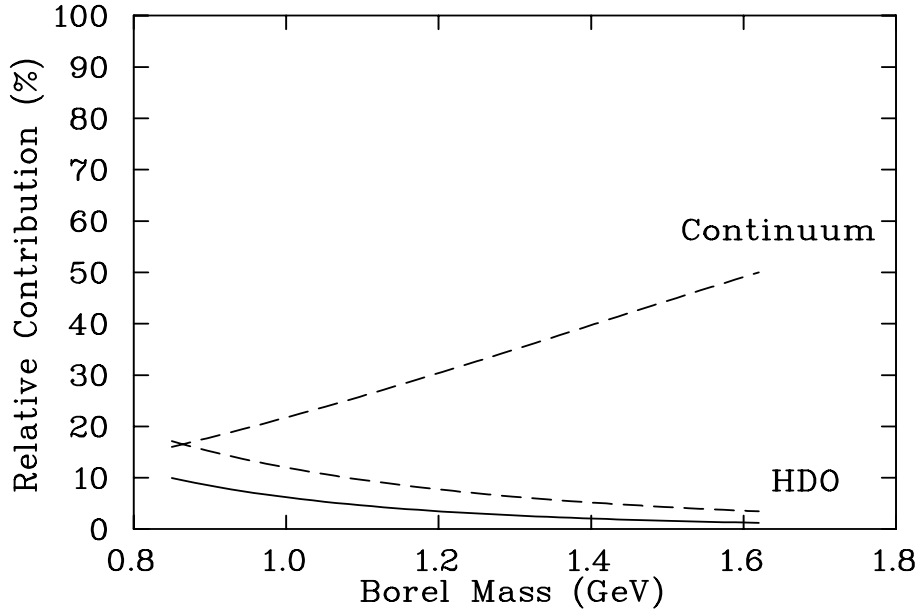,height=8cm,width=12cm,angle=90}}
\vspace{1cm}
\caption{The valid Borel window for sum rule~(\protect\ref{sum2}) at 
optimal $\beta=-1.2$.
Both the relative HDO contributions limited 
to 10\% of the OPE and the 
continuum-plus-transition contributions limited to 50\%
of the phenomenology are illustrated by dashed lines. 
Note that the lower limit is determined at $\beta=0$.}
\label{HIDCON8}
\end{figure}
%
The figure reveals that around a Borel mass of 1 GeV, the HDO
contributions are large at about 70\%, indicating poor OPE
convergence.  Only when the Borel mass reaches about 1.8 GeV, do their
contributions drop to below 10\%.  On the other hand, at such Borel
mass the continuum contributions increase to about 80\%, and almost
completely dominate
the phenomenology.  These continuum model contributions are subtracted
from the leading terms of the OPE such that the terms independent of
the Borel mass dominate the OPE.  This effect trivially explains why
in Ref.~\cite{Henley92} a plateau was reached when the Borel mass was
greater than 1.8 GeV.  As a result, any ground state properties
extracted from this sum rule are seriously contaminated by
short-comings of the excited state model.  We analyzed this sum rule
using the Monte-Carlo method with 1000 sets of QCD parameters, and the
conventional conditions: in the Borel region of 0.9 GeV to 1.2 GeV,
with the Ioffe current ($\beta=-1$), and the previous assumption that
the continuum threshold in the $g_A$ sum rule is the same as that of
the mass sum rule.  The result is $g_A=11.0\pm 4.86$, as compared to
1.26.  This example shows the importance of maintaining both OPE
convergence and ground state dominance in extracting hadron properties
from QCD sum rules.

Having demonstrated the failure of sum rule (\ref{sum2}), we now turn
to sum rule ~(\ref{sum3}) at the chirally-even structure
$iZ_\mu\sigma^{\mu\nu}p_\nu\gamma_5$.  This sum rule does have a valid
Borel window, as shown in Fig.~\ref{HIDCON8}. Within a wide region of
about 1 GeV, the HDO contributions to the OPE are less than 10\%, and
the continuum contributions are less than 50\%.  Analysis of this sum
rules yields $g_A=1.87\pm 0.92$, as explained below.

In illustrating how well a sum rule works, we first cast the  
sum rule into the subtracted form 
$\Pi_S= c\;g_A\, \tilde{\lambda}^2\;e^{-M_N^2/M^2}$ 
where $\Pi_S$ represents the OPE minus excited state contributions.
Here $c$ is a constant factor: $1$ for (\ref{sum2}),
(\ref{mix4}), (\ref{mix6}), (\ref{mix7});  $M_N$ for (\ref{sum1}), 
(\ref{sum3}), (\ref{mix1}), (\ref{mix2}), (\ref{mix3}), (\ref{mix5}); 
and $1/M_N$ for (\ref{mix8}).
Then the logarithm of both sides is plotted against the inverse of $M^2$.
The right-hand side will appear as a straight line.
The linearity of the left-hand side gives a good indication of 
OPE convergence and the quality of the continuum model.
The two curves should match for a good sum rule.
This way of matching the sum rules is similar to looking for a 
`plateau' as a function of Borel mass in the conventional analysis, 
but has the advantage of not restricting the analysis regime in Borel 
space to the valid regimes common to {\it both} two-point and 
three-point correlation functions.

\begin{figure}[p]
\centerline{\psfig{file=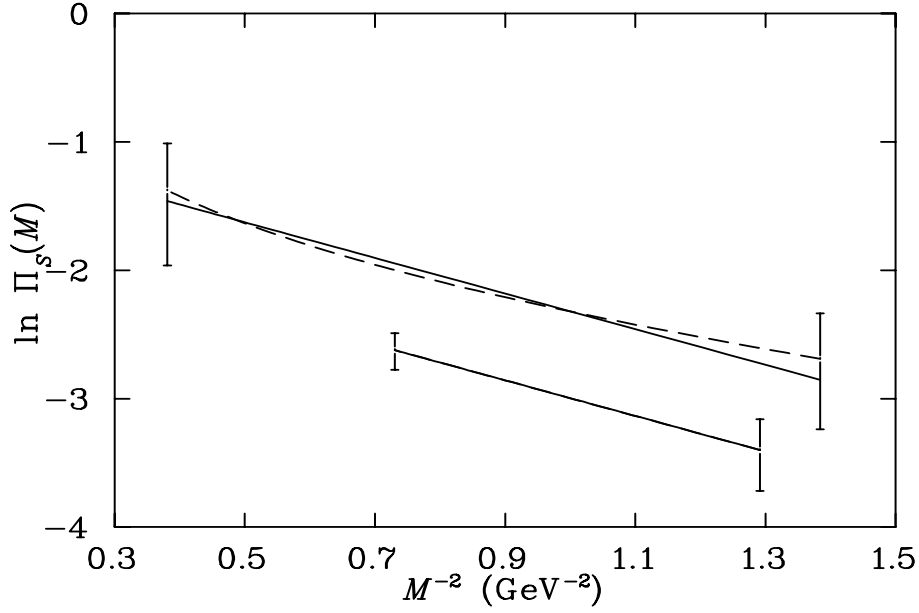,height=8cm,width=12cm,angle=90}}
\vspace{1cm}
\caption{A six-parameter ($M_N$, $\tilde{\lambda}_{1/2}$, 
$w_N$, $w$, $A$, $g_A$) fit of sum rule~(\protect\ref{sum3}) in 
conjunction with~(\protect\ref{nuc2}).
The solid line is the ground state contribution: 
The dashed line (hidden for~(\protect\ref{nuc2})) is
the rest of the contributions (OPE$-$continuum$-$transitions).
The error bars are only shown at the two ends for clarity.}
\label{RHSLHS8}
\end{figure}

\begin{figure}[p]
\centerline{\psfig{file=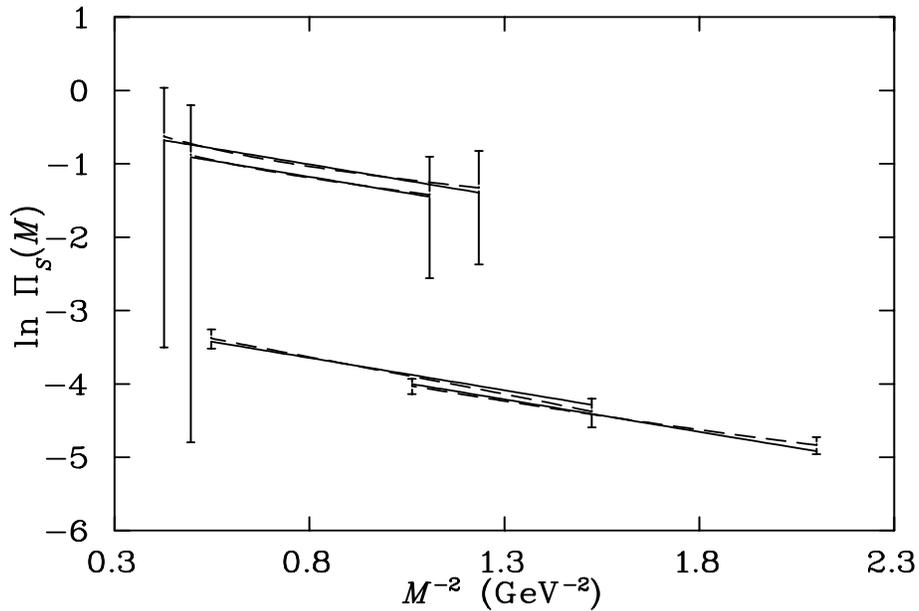,height=8cm,width=12cm,angle=90}}
\vspace{1cm}
\caption{Same as Fig.~\protect\ref{RHSLHS8}, except for 
the mixed correlator sum rules.  From top down, the fits are for sum
rules~(\protect\ref{mix2}), (\protect\ref{mix5}),
(\protect\ref{nucmix3}), and (\protect\ref{nucmix2}).  For clarity,
the mass sum rule fits are shifted downward by 2 units.}
\label{RHSLHS03}
\end{figure}

Fig.~\ref{RHSLHS8} shows the fit of ~(\ref{sum3}) in conjunction
with~(\ref{nuc2}) at the optimal $\beta=-1.2$.  Fig.~\ref{RHSLHS03}
shows the fits of ~(\ref{mix2}) and (\ref{mix5}) in conjunction
with~(\ref{nucmix2}) and ~(\ref{nucmix3}) at the optimal $\beta=0$.
For comparison purposes, the corresponding nucleon mass sum rule is
also plotted.

Sum rule~(\ref{sum3}) does not have enough information in the OPE to
completely determine the fit parameters, so we have assumed the
equivalence of continuum thresholds in two- and three-point functions;
$w=w_N$.  For curiosity, we also tried to fix the nucleon mass at its
known value and found the fits and uncertainties are essentially the
same.  There are some small deviations from linearity (dashed lines)
for the spin-1/2 correlator $g_A$ sum rules, but near perfect
linearity for the mixed correlator $g_A$ sum rules.  Unfortunately,
the error bars for the $g_A$ sum rules are much larger than those of
the corresponding nucleon mass sum rules.
%
%
%

\begin{figure}[p]
\centerline{\psfig{file=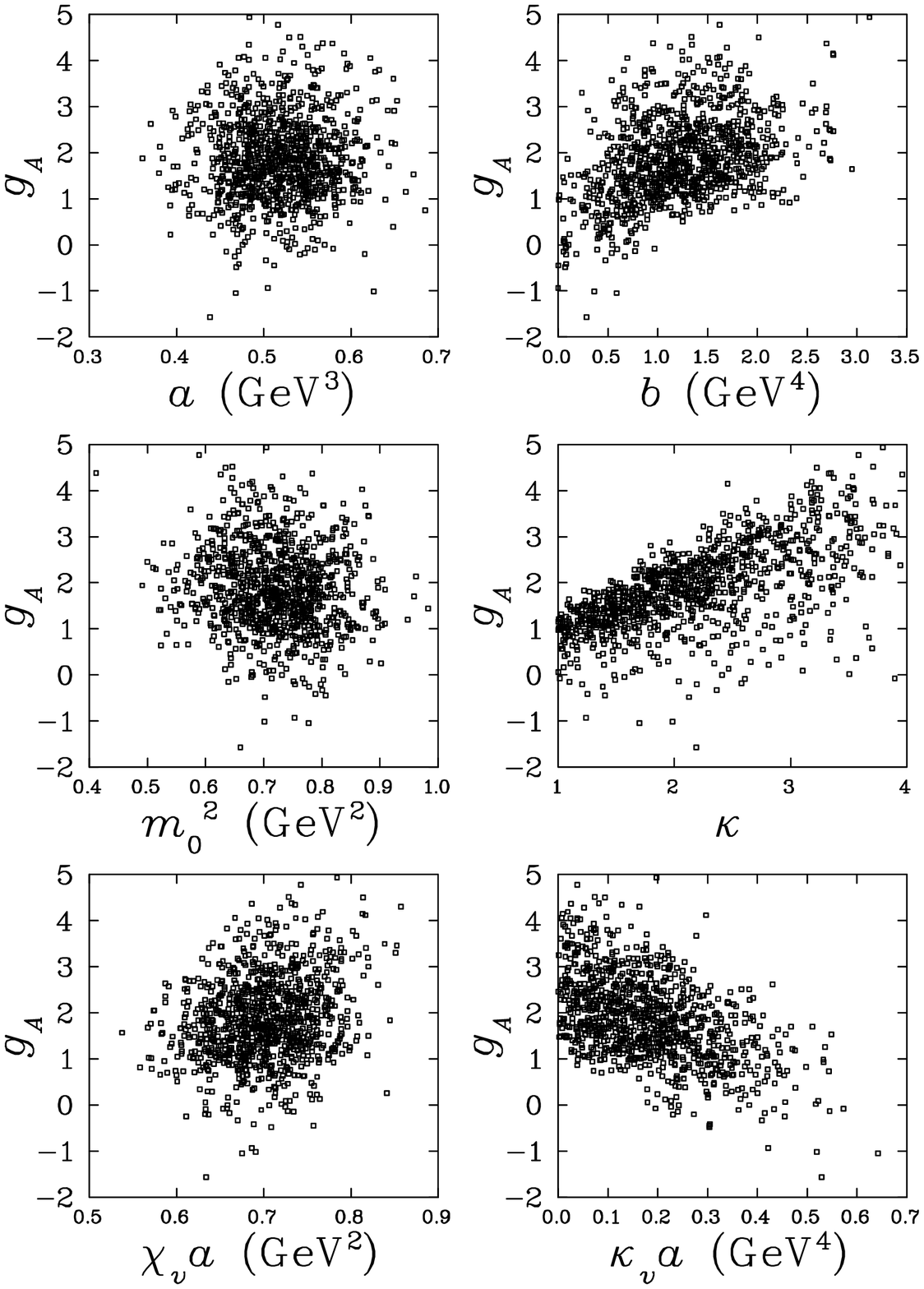,height=10cm,width=11cm}}
\vspace{1cm}
\caption{Scatter plots showing correlations between $g_A$ 
and the QCD input parameters for sum rule~(\protect\ref{sum3}).}
\label{CORRGAL8}
\end{figure}

\begin{figure}[p]
\centerline{\psfig{file=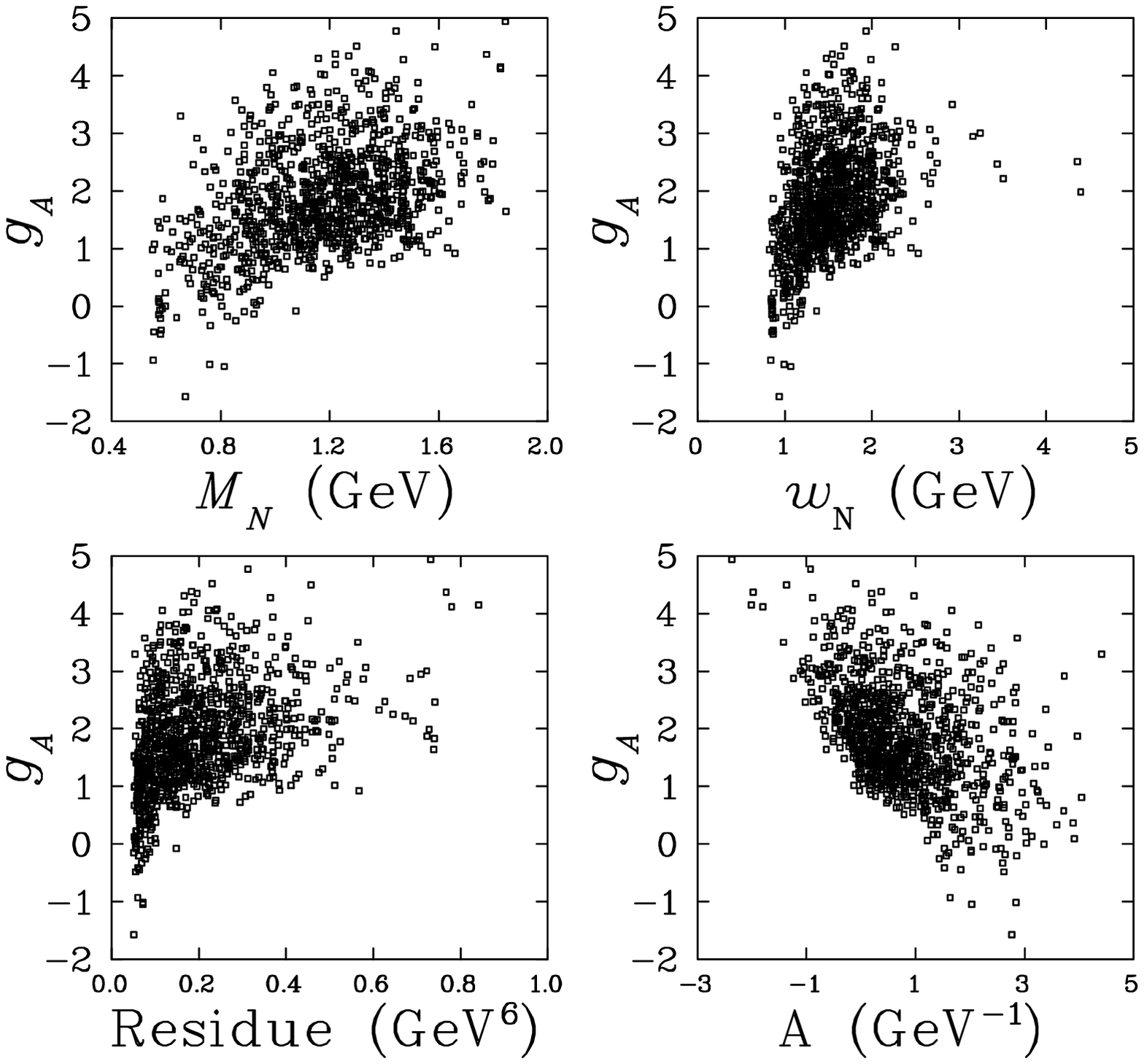,height=7cm,width=11cm}}
\vspace{1cm}
\caption{Scatter plots showing correlations between $g_A$ 
and the phenomenological fit parameters for sum rule~(\protect\ref{sum3}).}
\label{CORRGAR8}
\end{figure}

Since all the fit parameters in the Monte-Carlo analysis are
correlated, one can study the correlations between any two parameters
by looking at their scatter plots.  Fig.~\ref{CORRGAL8} shows the
correlations of the QCD input parameters with $g_A$ for sum
rule~(\ref{sum3}).  The plots for the quark condensate $a$, the mixed
condensate $m^2_0$ and the vacuum susceptibility $\chi_v a$ look
fairly random, suggesting little correlation.  The gluon condensate
and the factorization violation parameter display some weak positive
correlations with $g_A$, while the vacuum susceptibility $\kappa_v a$
reveals weak negative correlations.
In fact, we found the  same correlation patterns 
between $g_A$, $\kappa$ and $\kappa_v$ in all the sum rules.
Fig.~\ref{CORRGAR8} shows the correlations of the 
phenomenological fit parameters with $g_A$ for sum rule~(\ref{sum3}).

%
%

\begin{figure}[p]
\centerline{\psfig{file=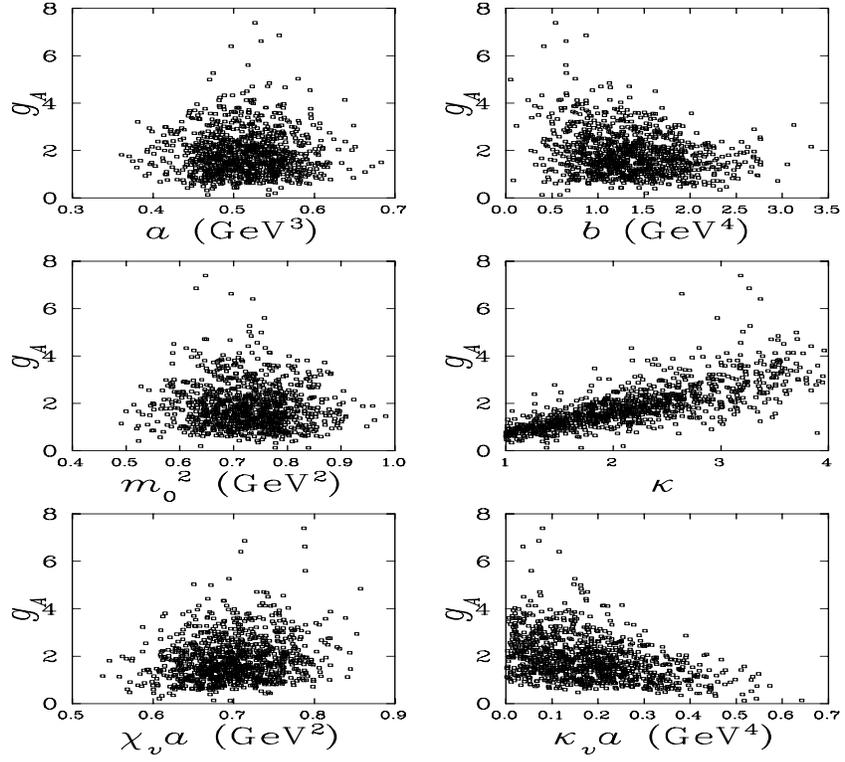,height=10cm,width=11cm}}
\vspace{1cm}
\caption{Scatter plots showing correlations between $g_A$ 
and the QCD input parameters for sum rule~(\protect\ref{mix5}).}
\label{CORRGAL13}
\end{figure}

\begin{figure}[p]
\centerline{\psfig{file=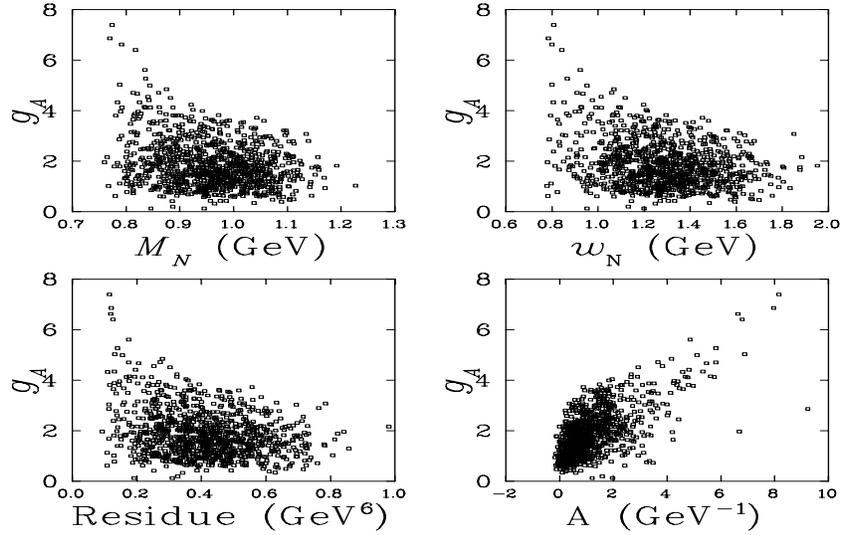,height=7cm,width=11cm}}
\vspace{1cm}
\caption{Scatter plots showing correlations between $g_A$ 
and the phenomenological fit parameters for sum
rule~(\protect\ref{mix5}).}
\label{CORRGAR13}
\end{figure}

One should bear in mind, however, that the correlations are in general
different from sum rule to sum rule.  Fig.~\ref{CORRGAL13} shows the
correlations of the QCD input parameters with $g_A$ for sum
rule~(\ref{mix5}).  Fig.~\ref{CORRGAR13} shows the correlations of the
phenomenological fit parameters with $g_A$ for sum rule~(\ref{mix5}).
There are some subtle differences between the correlations in this sum
rule and those in~(\ref{sum3}).  First, $g_A$ is predicted to be
positive here in the entire parameter space, unlike ~(\ref{sum3}).
Second, the distributions in Fig.~\ref{CORRGAL8} are rounded-shaped,
while they are more concentrated toward smaller $g_A$ in
Fig.~\ref{CORRGAL13}.  Third, the correlations with the gluon
condensate in~(\ref{mix5}) are weakly negative, opposite to those in
~(\ref{sum3}).  Fourth, the correlations among fit parameters are
quite different in the two sum rules.
It is interesting to see that in Fig.~\ref{CORRGAR13}, small values 
of the pole residue are responsible for large values of $g_A$  
and {\it vice versa}.

We have examined the scatter plots for other sum rules as well.  We
find that the above differences between the mixed correlator and the
spin-1/2 correlator are qualitatively true for other sum rules. It is
interesting to observe how different sum rules resolve the same
observables in different ways.  This reinforces the danger of linking
a particular term in the OPE as responsible for a particular
observable.
%
%

In Table~\ref{tabga} we give a summary of the results from the
consideration of 1000 sets of QCD parameters.  Only those sum rules
that have valid Borel windows are listed.  The presence of a second
row indicates that the continuum threshold was successfully searched
as an independent parameter (the second or third scenario discussed
earlier).  Note that sum rule~(\ref{mix5}) for $g^8_A$, $g^s_A$ and
$g^0_A$ has the continuum model term proportional to $\beta$.
Therefore $\beta=-0.1$ was used instead of $\beta=0$ in order to 
maintain a model for excited state contributions in place of additional
poles, and enhance the magnitude of the HDO of this sum rule.

The positive statement one can make is that for each coupling, the
results are consistent with each other within one standard deviation.
The results also agree with the experimental values within one and
half standard deviations, although the central values appear to lie
consistently above the experimental values.  Sum rule~(\ref{mix2})
consistently predicts larger values for the couplings than the rest.
Also the correct ordering $g^8_A>g^s_A>g^0_A$ is apparent.
Unfortunately, the uncertainties in the coupling constants are large
at about 50\% to 100\%, as compared to the nucleon mass obtained from
the same method, which only has 10\% to 25\% errors [see
Eq.(\ref{fit1}) and Eq.(\ref{fit2})].  The large uncertainties suggest
that fine adjustments of the QCD input parameters will allow one to
make the central values
of the fit parameters reproduce the experimental numbers.
At the present stage, we feel that such refinements are not meaningful
until the sum rules themselves are derived to a similar accuracy.

When the continuum threshold is included as a search parameter, the
only stable results for $g_A$ are from sum rule~(\ref{mix5}).  This
sum rule proves to be the best sum rule we derived.  The overall
effects of searching the continuum threshold appear to cause small
increases in the couplings, and small decreases in the transition
strengths.  However, the uncertainties are too large to allow a
definite conclusion.  We have compared the quality of the fits with or
without searching the continuum threshold and found they are
essentially the same.

The origin of the large errors lies mainly in the poorly determined
normalization of the two-point function as indicated by the large
uncertainty in $\tilde{\lambda}^2$.  While we can consider ratios of
two- and three-point functions to eliminate the parameter
$\tilde{\lambda}^2$, the poorly determined normalization of the
two-point function remains. The hope was that the two- and three-point
functions would be sufficiently correlated that the poorly determined
normalization of the two-point function would mirror that of the
three-point function and allow an extraction of $g_A$ with small
uncertainties. However, our analysis indicates this is not the case.

Some interplay between $g_A$ and $\tilde{\lambda}^2$ is illustrated in
Figs.~\ref{CORRGAR8} and in particular~\ref{CORRGAR13}.  The
correlations exist because the valid Borel regimes for the sum rules
are not identical, and the residue does not factor out as it would in
a ratio of sum rules. Fig.~\ref{CORRGAR13} displays the tendency for
$g_A$ to be small when $\tilde{\lambda}^2$ is large and {\it vice
versa}, as one might expect from the form $g_A\,
\tilde{\lambda}^2\;e^{-M_N^2/M^2}$ for the three-point function ground
state contribution.  We expect the result to be general to any QCD sum
rule of nucleon current matrix elements (three-point functions) from
the linear response of the external-field.  Only when the
external-field is treated non-perturbatively,
might the drawback be avoided~\cite{Burk96}.

Other sources of uncertainty are in the poorly known vacuum
susceptibility $\kappa_v$ and in the factorization approximation of
higher order operators.  From correlation studies we discovered that
the sum rules seem to favor smaller factorization violation and larger
vacuum susceptibility $\kappa_v$.  Better estimates of these
parameters are clearly needed.

The presence of the transitions is an additional source of
contamination not found in two-point functions.  To see the role the
transitions play, it is useful to cast the sum rule in the form
$g_A+AM^2=\cdots$. By matching the two sides over some region in $M^2$
where a linear behavior is observed, one can extract $g_A$ from the
intercept of the straight line with the y-axis.  In fact this is the
method explicitly used in previous analyses to extract $g_A$, except
that some combinations with the mass sum rule or other sum rules were
used.  It is clear that the transitions play a crucial role since they
determine the slope of the line from which $g_A$ is extracted.  Small
changes in the transition strength $A$ can lead to quite different
values of $g_A$.
%
\begin{table}[tb]
\caption{Monte-Carlo analysis results from consideration of 
1000 QCD parameter sets.  A second row indicates that the continuum
threshold could be searched as an independent parameter for that sum
rule.  Otherwise, it is assumed to be the same as that of the nucleon
mass sum rule.  The last column shows the relative contributions of
the continuum (including transitions) to the phenomenological side in
the given Borel window.}
\label{tabga}
\begin{tabular}{cccccc}
Sum Rule & Borel Window (GeV) & $g_A$ & A $(\mbox{GeV}^{-1})$ & $w$ (GeV)
& Continuum \\ \hline
(\protect\ref{sum3}) 
& 0.85 to 1.62 & 1.87$\pm$ 0.92 & 0.77$\pm$ 0.97 &     & 29\% to 50\% \\ 
(\protect\ref{mix2}) 
& 0.90 to 1.53 & 2.52$\pm$ 1.31 & 0.40$\pm$ 0.95 &   & 20\% to 50\% \\ 
& 0.90 to 1.55 & 3.53$\pm$ 1.44 &-0.72$\pm$ 0.43 & 3.49$\pm$4.73 
& 20\% to 50\% \\ 
(\protect\ref{mix5}) 
& 0.95 to 1.42 & 1.88$\pm$ 0.94 & 1.20$\pm$ 1.13 &  & 31\% to 50\% \\ 
& 0.95 to 2.34 & 2.38$\pm$ 1.07 & 0.96$\pm$ 0.92 & 2.08$\pm$0.42 
& 20\% to 50\% \\  \hline
Sum Rule & Borel Window (GeV) & $g^8_A$  
& A $(\mbox{GeV}^{-1})$ & $w$ (GeV) & Continuum \\ \hline
(\protect\ref{sum3}) 
& 0.67 to 1.16 & 2.26$\pm$ 2.28 & -1.49$\pm$ 2.42 &  & 22\% to 50\% \\ 
(\protect\ref{mix2}) 
& 0.71 to 1.60 & 2.55$\pm$ 1.10 & 0.13$\pm$ 0.86 &  & 0.6\% to 50\% \\ 
(\protect\ref{mix5}) 
& 1.10 to 1.23 & 1.43$\pm$ 0.84 & 1.28$\pm$ 1.06 &  & 46\% to 50\% \\ 
& 1.10 to 1.50 & 1.57$\pm$ 0.82 & 1.03$\pm$ 0.80 & 2.97$\pm$9.25 
& 36\% to 50\% \\  \hline
Sum Rule & Borel Window (GeV) & $g^s_A$  
& A $(\mbox{GeV}^{-1})$ & $w$ (GeV) & Continuum \\ \hline
(\protect\ref{mix2}) 
& 0.59 to 1.04 & 1.97$\pm$ 1.43 &-1.33$\pm$ 0.72 &  & 21\% to 50\% \\ 
(\protect\ref{mix5}) 
& 0.96 to 1.95 & 0.70$\pm$ 0.58 & 0.51$\pm$ 0.63 &  & 33\% to 50\% \\ 
& 0.96 to 1.96 & 1.06$\pm$ 0.51 & 0.51$\pm$ 0.41 & 3.72$\pm$5.40 
& 11\% to 50\% \\
(\protect\ref{mix8}) 
& 0.73 to 1.01 & 0.67$\pm$ 0.47 & 0.94$\pm$ 0.76 &     
& 34\% to 50\% \\  \hline
Sum Rule & Borel Window (GeV) & $g^0_A$ & A $(\mbox{GeV}^{-1})$ 
& $w$ (GeV) & Continuum \\ \hline
(\protect\ref{mix2}) 
& 0.64 to 0.70 & 0.81$\pm$ 0.99 &-1.36$\pm$ 1.20 &      & 45\% to 50\% \\ 
(\protect\ref{mix5}) 
& 0.77 to 1.70 & 0.39$\pm$ 0.30 &-0.12$\pm$ 0.22 &      & 12\% to 50\% \\ 
& 0.77 to 1.74 & 0.55$\pm$ 0.28 &-0.02$\pm$ 0.10 & 3.66$\pm$ 3.15  
& 12\% to 50\% \\ 
\end{tabular}
\end{table}

\section{Alternative Treatment of the Transitions}
\label{pc}

As discussed in Sec.~\ref{ansatz}, the transitions between the ground
state and the excited states lead to off-diagonal contributions that
are not exponentially suppressed relative to the ground state.  We
have seen in the previous section that these off-diagonal
contributions play important roles since they determine the curvature
from which $g_A$ is extracted.  As can be seen from
Eq.~(\ref{phen-gen-borel}), the contribution from the transitions is
in general a complicated function of the Borel mass. The usual
treatment of crudely modeling the transitions by a constant parameter
(multiplied by $e^{-M_N^2/M^2}$) could have significant impact on the
determination of $g_A$. Recently, a new formalism for treating the
transitions has been pointed out in Ref.~\cite{Jin96}.  In this new
formalism, the transition contribution is exponentially suppressed
relative to the ground state contribution and hence can be included in
the continuum model. Here we briefly illustrate the new formalism. The
reader is referred to Ref.~\cite{Jin96} for further details and
discussions.

Let us consider the phenomenological representation of Eq.\
(\ref{pole}).  If one first multiplies the expression with the factor
$(p^2-M_N^2)$, then performs the Borel transform, one sees that the
transition contributions are exponentially suppressed relative the
ground state contributions.  As a result, one can use the conventional
pole plus continuum model on the phenomenological side.  The physics
behind such a manipulation is a rearrangement of information: the
transition strength on the phenomenological side is not lost, but gets
absorbed into the new continuum model built from an altered OPE.  So
it is important in this new formalism to treat the continuum threshold
as an independent phenomenological parameter to be extracted from the
sum rule along with the ground state property of interest.  This point
has been emphasized in Ref.~\cite{Jin96}, where an example was given
in the scalar channel. (This point, however, was completely ignored in
Refs.~\cite{Balitsky83,Braun87,HeJi96}, where a similar formalism was
adopted).

The sum rules in this new formalism can be obtained from the standard ones 
by the following substitutions.
On the OPE side, 
$M^6\rightarrow 3M^8-M_N^2 M^6$,
$M^4\rightarrow 2M^6-M_N^2 M^4$,
$M^2\rightarrow  M^4-M_N^2 M^2$,
$M^0\rightarrow     -M_N^2    $,
$1/M^2\rightarrow  -(1+M_N^2/M^2)$,
$1/M^4\rightarrow  -(1/M^2+M_N^2/2M^4)$.
The $E$ factors associated with powers of the Borel mass 
are suppressed for clarity.
On the phenomenological side, for the spin-1/2 correlator, one
replaces the right hand side of Eq.~(\ref{sum1}) to Eq.~(\ref{sum3})
by $c\,\tilde{\lambda}^2_{1/2}\;g_A\;e^{-M_N^2/M^2}$, where $c$ is
$2M_N^2$, $-1$, $-M_N$, respectively.  Similarly the phenomenological
side of the mixed correlator becomes
$c\,\tilde{\lambda}_{1/2}\tilde{\lambda}_{3/2}\;g_A\;e^{-M_N^2/M^2}$,
where $c$ is $2M_N^2$, $2M_N$, $-M_N$, $-1$,
$-M_N$,$-1$,$-1$,$-1/M_N$, for Eq.~(\ref{mix1}) to Eq.~(\ref{mix8}),
respectively.  We will denote these modified sum rules by appending a
symbol (PC) indicating pole$+$continuum phenomenology to their
standard counterparts.

In Table~\ref{tabgaPC} we summarize all the results from a Monte-Carlo
analysis of 1000 QCD parameter sets using the new formalism.  For
comparison purposes, we also list the results from not searching the
continuum threshold.

The results for the coupling constants are qualitatively the same as
those in the standard approach.  This suggests that the main source of
error in $g_A$ is not in the treatment of transitions.  However, the
new formalism does show the potential for improvement.  For example,
the Borel windows become generally wider, the continuum contributions
become smaller, and the numerical fits are generally more stable.
Future studies should take advantage of the new formalism to avoid the
errors associated with the approximation in modeling the unwanted
transitions.

%
\begin{table}[tb]
\caption{Same as Table~\protect\ref{tabga}, except they are obtained 
from the new pole-plus-continuum sum rules, as indicated by (PC).}
\label{tabgaPC}
\begin{tabular}{ccccc}
Sum Rule & Borel Window (GeV) & $g_A$ & $w$ (GeV) & Continuum \\ \hline
(\protect\ref{sum3})(PC) 
& 0.82 to 1.83 & 1.44$\pm$ 0.59 &                & 2.3\% to 50\% \\ 
(\protect\ref{mix2})(PC) 
& 0.84 to 1.40 & 2.80$\pm$ 1.50 &                & 9\% to 50\% \\ 
& 0.84 to 1.65 & 3.60$\pm$ 1.76 & 2.22$\pm$ 0.62 & 0.5\% to 50\% \\ 
(\protect\ref{mix5})(PC) 
& 0.95 to 1.60 & 2.02$\pm$ 1.02 &                & 9.3\% to 50\% \\ 
& 0.95 to 1.81 & 2.44$\pm$ 1.12 & 2.15$\pm$ 0.48 & 10\% to 50\% \\ \hline
Sum Rule & Borel Window (GeV) & $g^8_A$ & $w$ (GeV) & Continuum \\ \hline
(\protect\ref{sum3})(PC) 
& 0.65 to 1.36 & 1.40$\pm$ 1.03 &                & 2\% to 50\% \\ 
& 0.65 to 1.65 & 1.75$\pm$ 1.22 & 2.48$\pm$ 2.08 & 0\% to 50\% \\ 
(\protect\ref{mix2})(PC) 
& 0.67 to 1.48 & 2.44$\pm$ 1.30 &                & 2\% to 50\% \\ 
& 0.67 to 1.77 & 3.19$\pm$ 1.54 & 2.39$\pm$ 0.57 & 0\% to 50\% \\ 
(\protect\ref{mix5})(PC) 
& 1.13 to 1.96 & 1.43$\pm$ 0.75 &                & 9\% to 50\% \\ 
& 1.13 to 2.17 & 1.65$\pm$ 0.84 & 2.07$\pm$ 0.78 & 9\% to 50\% \\  \hline
Sum Rule & Borel Window (GeV) & $g^s_A$ & $w$ (GeV) & Continuum \\ \hline
(\protect\ref{sum3})(PC) 
& 0.52 to 1.09 & 0.62$\pm$ 1.48 &                & 1\% to 50\% \\ 
& 0.52 to 1.33 & 0.54$\pm$ 0.99 & 2.16$\pm$ 2.26 & 0\% to 50\% \\ 
(\protect\ref{mix2})(PC) 
& 0.57 to 1.24 & 1.20$\pm$ 0.94 &                & 2\% to 50\% \\ 
& 0.57 to 1.50 & 1.56$\pm$ 1.08 & 2.39$\pm$ 2.15 & 0\% to 50\% \\ 
(\protect\ref{mix5})(PC) 
& 0.98 to 1.68 & 0.79$\pm$ 0.63 &                & 9\% to 50\% \\ 
& 0.98 to 1.90 & 0.97$\pm$ 0.62 & 2.30$\pm$ 0.97 & 1\% to 50\% \\  \hline
Sum Rule & Borel Window (GeV) & $g^0_A$ & $w$ (GeV) & Continuum \\ \hline
(\protect\ref{sum3})(PC) 
& 0.45 to 1.03 & -0.36$\pm$ 0.90 &                & 0.3\% to 50\% \\ 
(\protect\ref{mix2})(PC) 
& 0.60 to 0.91 & 0.36$\pm$ 0.59 &                & 9\% to 50\% \\ 
& 0.60 to 1.11 & 0.56$\pm$ 0.59 & 2.22$\pm$ 1.79 & 0.5\% to 50\% \\   
(\protect\ref{mix5})(PC) 
& 0.76 to 1.44 & 0.42$\pm$ 0.32 &                & 5\% to 50\% \\ 
& 0.76 to 1.68 & 0.51$\pm$ 0.36 & 2.16$\pm$ 1.27 & 5\% to 50\% \\ 
\end{tabular}
\end{table}
%
%
\begin{figure}[p]
\centerline{\psfig{file=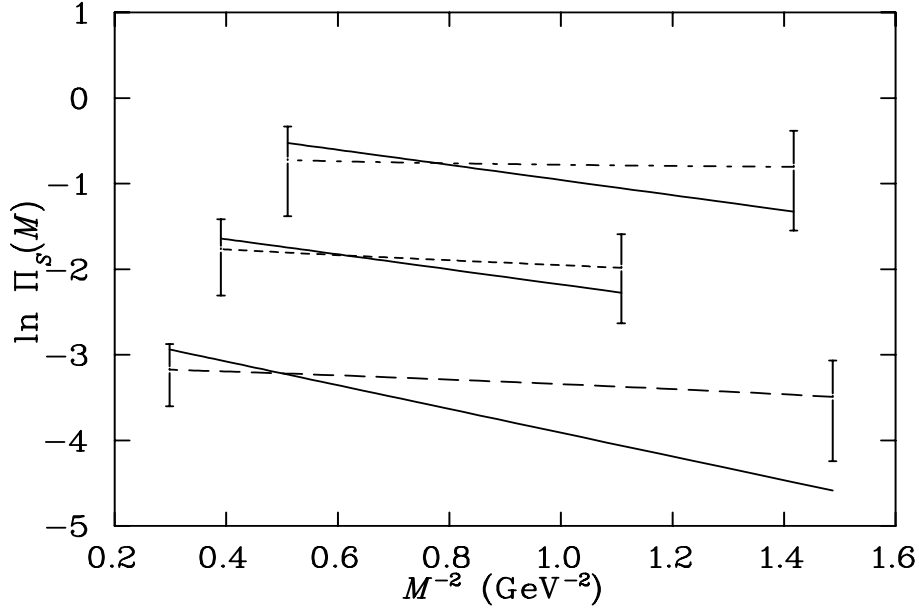,height=8cm,width=12cm,angle=90}}
\vspace{1cm}
\caption{From top down, Monte-Carlo based fits for $g_A$ from 
sum rules~(\protect\ref{mix2})(PC), (\protect\ref{mix5})(PC),
and~(\protect\ref{sum3})(PC).  The continuum thresholds are assumed
the same as that of the corresponding mass sum rules.  The solid line
is the ground state contribution, and the dashed line is the
OPE minus continuum.  For clarity, the fits
for~(\protect\ref{mix5})(PC) and (\protect\ref{sum3})(PC) are shifted
downward by 1 unit, respectively.}
\label{RHSLHS803PC}
\end{figure}
%
\begin{figure}[p]
\centerline{\psfig{file=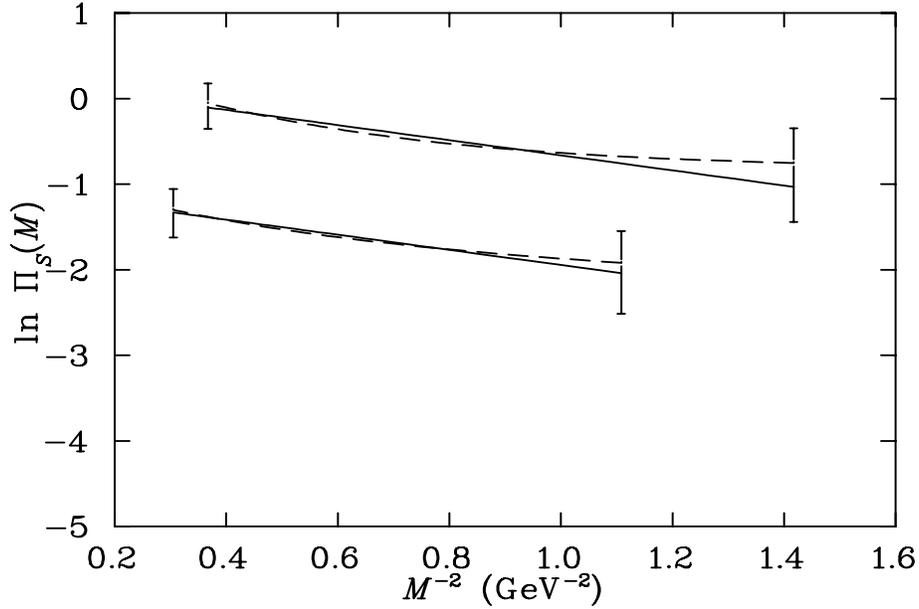,height=8cm,width=12cm,angle=90}}
\vspace{1cm}
\caption{From top down, fits of sum rules~(\protect\ref{mix2})(PC) and
(\protect\ref{mix5})(PC) in which the continuum threshold was
searched successfully.}
\label{RHSLHS803PCW}
\end{figure}
%

Although the extracted couplings appear to agree within uncertainties
with or without searching the continuum thresholds, there are
qualitative differences between the two.  To demonstrate this, we show
in Fig.~\ref{RHSLHS803PC} the fits for sum
rule~(\ref{sum3})(PC),~(\ref{mix2})(PC) and~(\ref{mix5})(PC), without
searching the continuum threshold.  In Fig.~\ref{RHSLHS803PCW} we show
the two successful fits obtained when searching the continuum
threshold.  We see that the two sides of the sum rules do not match
very well when the continuum threshold is not searched.  The match
improves significantly for the mixed correlator sum rules when the
continuum threshold is searched.  A large uncertainty was found for
the searched continuum threshold in sum rule~(\ref{sum3})(PC). This
suggests that it is really unable to separate the pole from the
continuum.

The above results confirm that the continuum threshold in the new
formalism is relied upon to compensate for the transition strength
that was explicitly taken into account in the standard approach.
Fixing it in the new approach would introduce an artificial bias to
the extracted ground state spectral properties.  This is in contrast
with the situation in the standard approach.  There the OPE was not
altered and the transitions were modeled explicitly. As a result, the
quality of the fits with or without searching the continuum thresholds
was essentially the same.

\section{Summary and Conclusions}
\label{con}

We have derived eleven new QCD sum rules for the nucleon axial vector
coupling constants using the external-field method and generalized
interpolating fields.  Three are from the spin-1/2 correlator, which
can be reduced and compared to those in the literature.  Eight are
from the mixed correlator, which are new.  Using the Monte-Carlo
analysis, we are able to determine the predicative ability of the new
sum rules for $g_A$ for the first time.

The main advantage of the Monte-Carlo analysis is that it takes into
account all uncertainties in the QCD input parameters simultaneously,
allowing a quantitative study of how the uncertainties in the QCD
input parameters propagate to the phenomenological fit parameters.
The method carefully monitors the OPE convergence and the ground-state
dominance, the two key criteria in order for the QCD sum rule method
to work.  Together they determine the Borel window over which the two
sides of the sum rules are matched.  Those sum rules which do not have
a valid Borel window are considered unreliable and therefore
discarded.  Conventional QCD sum rule analysis is limited to only a
small portion of the QCD parameter space, and often the uncertainties
assigned to the fit parameters are not based on rigorous error
analysis but rather have a certain degree of arbitrariness.

Our most important findings in this work are: a) The nucleon axial
vector coupling constants calculated from standard QCD sum rule method
have large uncertainties associated with them, approximately 50\% to
100\%, as compared to the nucleon mass obtained from the same method
which has only 10\% to 25\% error.  Within uncertainties, the numbers
obtained from different sum rules are consistent with each other and
with the experimental values at 1.5$\sigma$.  The correct ordering in
magnitude of the coupling constants, $g^8_A>g^s_A>g^0_A$, is also
predicted by the sum rules.  b) Sum rule (\ref{sum2}), upon which the
previous analyses of $g_A$ are based, has poor OPE convergence and
poor ground-state dominance properties. The results extracted from
this sum rule are unreliable.  Both of these findings contradict the
conventional wisdom.  Traditionally, 10\% to 20\% errors are often
claimed for $g_A$ from QCD sum rule calculations without rigorous
error analysis.  The selection of sum rule (\ref{sum2}) was based upon
similarities with the mass sum rule~\cite{Chiu85} at the structure
$\hat{p}$, which was shown to have poor convergence
properties~\cite{Derek96}, or dimension arguments~\cite{Bely84}.  Our
Monte-Carlo analysis has shown that such criteria may not be reliable
in selecting sum rules.

The origin of the large errors in $g_A$ is three-fold.  First, it is
in the poorly determined nucleon mass sum rules, which are used to
normalize the couplings extracted from the form $g_A\,
\tilde{\lambda}^2\;e^{-M_N^2/M^2}$.
Second, two new parameters are needed to describe the response of the
QCD vacuum to the external field. These new parameters introduce
additional uncertainties, especially the vacuum susceptibility
$\kappa_v$ which is very poorly known.  Third, the transitions between
the ground state and the excited states caused by the probing axial
current is another source of uncertainty.  Their contributions are not
exponentially suppressed relative to the ground state and must be
included in the spectral representation.  Little is known about these
transitions. The standard approach is to introduce a new
phenomenological parameter to account for all the contributions from
the transitions.  Usually the new parameter is assumed to be constant,
but in fact it depends in some complicated way on the Borel mass.  The
approximation may have a sizable impact on the results since it plays
a crucial role in the extraction of $g_A$. In any event, the presence
of an additional unknown parameter leads to larger uncertainties in
$g_A$.

To further investigate the role of the transitions, we have studied 
an alternative treatment which can lead to exponential suppression 
of their contributions relative to the ground state.
As a result, one can apply the traditional pole-plus-continuum model 
in the phenomenological representation.
%
%
The contributions of the transitions are not lost, but simply get
absorbed in the new continuum model.  So it is important in this
formalism to search the continuum threshold as an independent
parameter.  The results for the coupling constants are similar to
those in the standard approach. This implies that the main source of
error is not in the treatment of transitions, but in the pole residues
and the vacuum susceptibilities. However, there are advantages to be
gained with the new formalism. The Borel windows become generally
wider,
and the numerical fits are generally more stable.  Future studies
should take advantage of the method to avoid errors associated with
the approximation of modeling of the unwanted transitions.

As for the question of valid sum rules for $g_A$, our Monte-Carlo
analysis reveals that only three out of the eleven sum rules are able
to resolve the ground state properties from the continuum. They are
sum rule (\ref{sum3}) from the spin-1/2 correlator, (\ref{mix2}) and
(\ref{mix5}) from the mixed correlator.  Even the three valid sum
rules perform differently.  Sum rule (\ref{mix5}) has the best
performance in terms of allowing a full search and the stability of
the results.  It is followed by sum rule (\ref{mix2}), then by
(\ref{sum3}).  We find that in general the mixed correlator sum rules
perform better than the spin-1/2 correlator sum rules, similar to the
situation for the nucleon mass sum rules.

Study of correlations among the input and fit parameters reveals how a
particular sum rule resolves the ground state from the continuum.  We
find that $g_A$ does not have strong correlations with the quark
condensate, the mixed condensate, the gluon condensate, nor the vacuum
susceptibility $\chi_v$.
$g_A$ has some weak positive correlations with the factorization
violation parameter $\kappa$, and some weak negative correlations with
the vacuum susceptibility $\kappa_v$.  These correlations can give
some hints on what values for $\kappa$ and $\kappa_v$ are preferred by
the self-consistency requirement in the sum rules. We should stress
that correlation patterns are different for different sum rules. The
above observations are some general trends displayed by all the valid
sum rules, despite some subtle differences.

As to the problem of how to reduce the large uncertainties in $g_A$
from QCD sum rule calculations, one may attack it from several
directions as suggested from the discussion of their origins: a)
improved accuracy of the QCD sum rules, b) better estimates of the
vacuum susceptibilities, especially $\kappa_v$, and c) better
treatment of the transitions.  Of these, point a) is the most
important.  One way to improve the accuracy of the sum rules is to
reduce the uncertainties in the QCD input parameters.  However, there
is a limit on it.  As shown in Ref.~\cite{Derek96}, the best one can
do with current implementation of the QCD sum rule approach is 25\%
for the two-point function normalization.  Reduction of this
uncertainty to the 10\% level requires 5\% uncertainties on the QCD
input parameters. But this level of accuracy is beyond the current
state of the art as indicated by the unacceptably large
$\chi^2/N_{\mbox{DF}}$.  We believe the future lies in improving the
sum rules themselves.  Examples may include incorporating $\alpha_s$
corrections, higher order power corrections, and devising alternative
interpolating fields that result in better sum rules, as done here.

Finally, we want to point out that although this work is focused on
$g_A$, the conclusions are expected to apply to other nucleon current
matrix elements.  The presence of large uncertainties in nucleon
matrix elements indicate that only small adjustments of the QCD input
parameters are required to reproduce the measured values. Likewise,
small fluctuations in the input parameters can lead to large
fluctuations in the matrix elements, such that most QCD sum rule
calculations of matrix elements in the literature are likely to have
50\% uncertainties at best, given the current implementation of the
QCD sum rule formalism.

\acknowledgements

This work is supported in part by the Natural Sciences and
Engineering Research Council of Canada and U.S. DOE under
grants DE-FG03-93DR-40774 (F.L.), DE-FG06-88ER-40427 (D.L.)
and DF-FC02-94ER-40818 (X.J.).



\end{document}